\documentclass{article}

\usepackage[english]{babel}

\usepackage[letterpaper,top=2cm,bottom=2cm,left=2cm,right=2cm,marginparwidth=1.75cm]{geometry}

\usepackage{amsmath}
\usepackage{amssymb}
\usepackage{authblk}

\usepackage{graphicx}
\usepackage{subcaption}
\usepackage{bm}
\usepackage{float}
\usepackage{array}
\usepackage[section]{placeins}
\usepackage{multirow}
\usepackage{multicol}
\usepackage[round]{natbib}
\usepackage{makecell}
\usepackage[colorlinks=true, linkcolor=blue, citecolor=blue, urlcolor=blue]{hyperref}

\begin{document}

%\title{Enhancing Wind Speed Predictions with Bias-Corrected Crowdsourced Data and Spatio-Temporal models}
\title{Enhancing the Accuracy of Spatio-Temporal Models for Wind Speed Prediction by Incorporating Bias-Corrected Crowdsourced Data}
\author[1]{Eamonn Organ\thanks{Corresponding author: \texttt{Eamonn.Organ@ul.ie}}}
\author[1]{Maeve Upton}
\author[2]{Denis Allard}
\author[2]{Lionel Benoit}
\author[1]{James Sweeney}

\affil[1]{Department of Mathematics and Statistics, University of Limerick, Ireland}
\affil[2]{Biostatistics and Spatial Processes (BioSP), INRAE, 84914 Avignon CEDEX 9, France}
\maketitle

\begin{center}
\small
This is the Author Accepted Manuscript of the article:
Organ, E., Upton, M., Allard, D., Benoit, L., and Sweeney, J. (2025),
``Enhancing the Accuracy of Spatio-Temporal Models for Wind Speed Prediction by Incorporating Bias-Corrected Crowdsourced Data,'' 
\textit{Environmetrics}.
The final Version of Record is available at:\\
\url{https://doi.org/10.1002/env.70069}. \\
This manuscript is made available under the Creative Commons Attribution 4.0 International (CC BY 4.0) licence:
\url{https://creativecommons.org/licenses/by/4.0/}.
\end{center}

\begin{abstract}
Accurate high-resolution spatial and temporal wind speed data is critical for estimating the wind energy potential of a location. For real-time wind speed prediction, statistical models typically depend on high-quality (near) real-time data from official meteorological stations to improve forecasting accuracy. Personal weather stations (PWS) offer an additional source of real-time data and broader spatial coverage than official stations. However, they are not subject to rigorous quality control and may exhibit bias or measurement errors.
This paper presents a framework for incorporating PWS data into statistical models for validated official meteorological station data via a two-stage approach. First, bias correction is performed on PWS wind speed data using reanalysis data. Second, we implement a Bayesian hierarchical spatio-temporal model that accounts for varying measurement error in the PWS data. This enables wind speed prediction across a target area, and is particularly beneficial for improving predictions in regions sparse in official monitoring stations. Our results show that including bias-corrected PWS data improves prediction accuracy compared to using meteorological station data alone, with a 5\% reduction in prediction error on average across all sites. The results are comparable with popular reanalysis products, but unlike these numerical weather models our approach is available in real-time and offers improved uncertainty quantification.

\end{abstract}

\section{Introduction}
In this paper we investigate the benefits of incorporating crowdsourced wind data — a previously untapped source of information — into spatio-temporal models for predicting wind speeds at unobserved locations in Ireland. While the focus in this paper is the use of wind speed data for the evaluation of the potential renewable energy resource at a specific location, as well as management of installed infrastructure, accurate data on wind speeds at precise temporal and spatial scales is crucial in a number of application areas. The include: tracking storm conditions \citep{lagerquist2017machine}, the spread of wild fires~\citep{beer1991interaction}, and modelling air pollution~\citep{sahoo2023estimating}. 
To combat the effects of climate change, improve air quality and achieve energy independence, EU countries aim to accelerate the transition to renewable electricity generation~\citep{EUTarget}. The EU has set a target of 45\% of electricity  production to be generated from renewable sources by 2030~\citep{EUTarget}. Similarly, the Irish government's Climate Action Plan 2024 (CAP24,~\citet{GovCAP2024}) has set a target of 80\% of national electricity production to be generated from renewable sources by 2030. Wind power is Ireland's primary renewable resource, contributing 35\% of Irish electricity in 2023 \citep{WindEnergyIreland2023}. In CAP24 the Irish government has set a target of 9GW of onshore wind energy by 2030, and at least 5GW of off-shore wind energy ~\citep{GovCAP2024}. Currently, the Republic of Ireland has approximately 4.8GW of onshore wind energy installed, highlighting a 4.2GW gap~\citep{WindEnergyIrelandLatest}. The peak daily electricity demand ranges between 4.5 and 5.5 GW, with highs of more than 6 GW during cold weather.

One of the main challenges inhibiting the transition to renewable energy is its highly variable nature~\citep{morales2013integrating}. Wind speeds exhibit significant spatial and temporal variability, particularly in a country such as Ireland, where the climate is heavily influenced by its proximity to the Atlantic Ocean \citep{fusco2010variability}. Substantial amounts of planning and analysis must therefore be performed before a new wind farm can be constructed, posing challenges of cost and development time. Furthermore, the intermittent nature of wind poses challenges for day-to-day grid management, as gaps between wind energy production and demand must be filled by traditional demand-response generators, such as gas and oil, or heavily polluting fossil fuels such as coal \citep{dowell2014short}.

Statistical models have played an important role in quantifying and modelling the uncertainty of wind speeds and associated wind power generation. Two of the primary applications of statistical models in wind energy are wind resource estimation at potential wind farm sites~\citep{murthy2017comprehensive} and forecasting wind speeds at future time points to improve the management of electricity grid~\citep{foley2012current}. While long-term (beyond 6 hours) forecasts rely on numerical weather prediction (NWP) models, forecasts on shorter timescales (sometimes called nowcasting) benefit from statistical models or hybrid models that combine numerical outputs with statistical techniques~\citep{foley2012current}. For statistical methods, real-time wind observations are required as inputs~\citep{sweeney2013reducing, sweeney2020future}. These observations typically come from official weather stations maintained by meteorological organizations, which are often spatially sparse~\citep{sweeney2013reducing, sweeney2020future}. For example, the Republic of Ireland has only 23 such stations, which cover an area of over $70,000\text{km}^{2}$, equating to approximately one station per $3,000\text{km}^{2}$. Other potential sources of wind speed data include satellite derived remote sensing measurements, however these are subject to biases and are primarily designed for offshore wind speed estimation~\citep{de2019validation}.

The novelty of our approach is to explore the incorporation of previously unstudied crowdsourced weather observations to increase the quantity and spatial resolution of real-time wind speed data in the Republic of Ireland. We focus on data sources available in near real-time, with the goal of improving operational capabilities and laying the groundwork for future research in nowcasting techniques. Crowdsourcing for data collection has been widely used in other fields, such as astronomy \citep{raddick2009galaxy}, ecology \citep{mengersen2017modelling}, and air pollution monitoring \citep{bonas2023calibration}. In the context of weather, crowdsourced data refers to individuals uploading weather measurements using personal weather stations (PWS), which has gained prominence with the advent of lower-cost sensors and internet connectivity. One such project is the weather observation website (WOW,~\citet{WowMap}) in collaboration with Ireland's official meteorological organisation, Met Éireann and other national meteorological services. There is extensive information available for the Republic of Ireland, the United Kingdom and the Netherlands at \href{https://wow.met.ie/}{wow.met.ie}. In these countries, there are more crowdsourced stations than official meteorological ones. For example, approximately 100 personal stations have uploaded data to the WOW website in the Republic of Ireland, and over 10,000 are available globally. This rich dataset has the potential to enhance high-resolution wind resource estimation and improve short-term forecasting. However, the lack of quality control for these data sources, along with possible biases and noisy observations often impedes their practical use. This is a common issue in crowdsourced data across all fields \citep{aceves2017accuracy}. For weather there has been a significant body of research on utilising crowdsourced data with potential quality issues ~\citep{chakraborty2020statistical,chen2018trust}, but the majority has been on variables such as temperature or air pressure, which demonstrate significantly less local variability than wind. 

Wind data presents unique challenges due to its high variability and dependence on localized factors such as obstacles and the altitude of anemometers (wind speed measurement instruments). Some studies have begun addressing these challenges. For example, \citet{droste2020assessing} analysed crowdsourced wind data in the Netherlands using descriptive statistics to examine the distribution of wind speeds at meteorological and PWS, in particular highlighting issues with devices underestimating true wind speeds. They applied a linear correction method based on co-located meteorological and PWS~\citep{droste2020assessing}. While this improves the accuracy of the crowdsourced stations, in our dataset, meteorological and crowdsourced stations are not co-located. ~\citet{chen2021quality} filtered crowdsourced stations by applying initial quality control measures such as examining correlation of the crowdsourced station with the nearest meteorological station, and ensuring temporal consistency with the measurements. With the remaining stations, bias correction was performed using a quantile mapping approach, with quantiles interpolated from meteorological stations. However, Ireland contains both low number of meteorological weather stations and a climate with highly variable wind speeds \citep{MetEireannWind}, which makes quantile interpolation uncertain.

In this paper, we propose two contributions to address these challenges. First, we introduce a novel bias correction method that uses physical weather models to predict wind speed distributions at the locations of personal weather stations. Specifically, we leverage reanalysis datasets, which combine historical observations with numerical weather models, to obtain an estimated distribution of wind speeds at each site. Raw wind speed measurements from PWS are then transformed to match this predicted distribution using a quantile transform.

Second, we extend existing spatial models by developing, to our knowledge, the first statistically rigorous framework that incorporates crowdsourced wind speed data, allowing for varying measurement error across different sources of observations. By incorporating crowdsourced observations from poorly gauged locations, we demonstrate a significant reduction in prediction error at unobserved sites.

This paper is structured as follows: in Section~\ref{SectionData} an overview of the datasets used is provided and quality control measures are discussed. The numerical weather models used for bias correction are also introduced. In Section~\ref{SectionBiasCorrect}, a bias correction is performed for the PWS data using quantile transformations and incorporating reanalysis data. In Section~\ref{SectionStat} spatial and spatio-temporal models are introduced along with extensions to mitigate noisy data. In Section~\ref{SectionResults} the models are applied to Irish wind data, and the predictive improvements of incorporating crowdsourced data are evidenced.

\section{Data}
\label{SectionData}

The wind data sources in this article consist of official meteorological data from Met Éireann, citizen science or crowdsourced wind data in the form of personal weather stations, and reanalysis data from numerical weather simulation models. We exclusively focus on data from the Republic of Ireland, an island located in the east Atlantic. Its climate is strongly influenced by the ocean, with prevailing southwesterly to westerly winds, and wind strongest in coastal locations \citep{MetEireannWind}.
An overview of each data source is provided in the following.
\subsection{Met Éireann Data}
\label{MetData}
\begin{figure}[!htbp]
    \centering

    \begin{subfigure}{0.45\textwidth}
        \includegraphics[width=\linewidth]{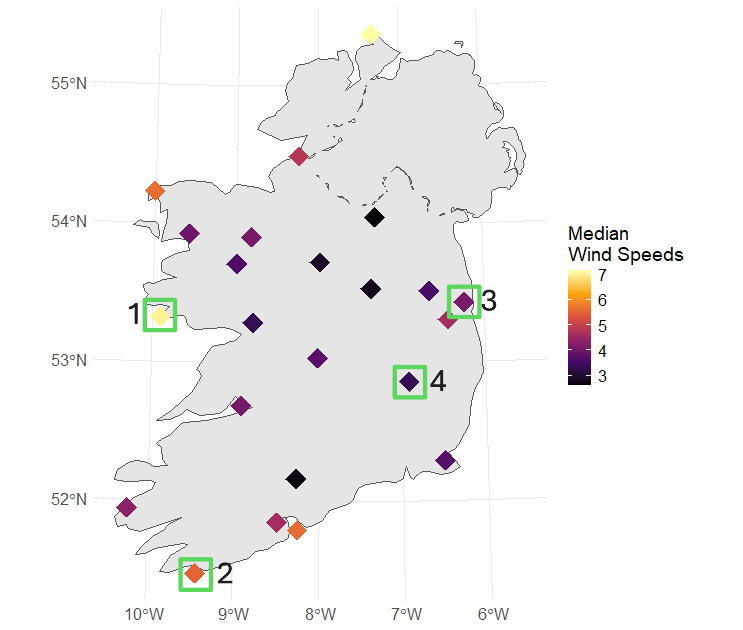}
        \caption{(a)}
        \label{fig:metmedian}
    \end{subfigure}
    \hfill
    \begin{subfigure}{0.45\textwidth}
        \includegraphics[width=\linewidth]{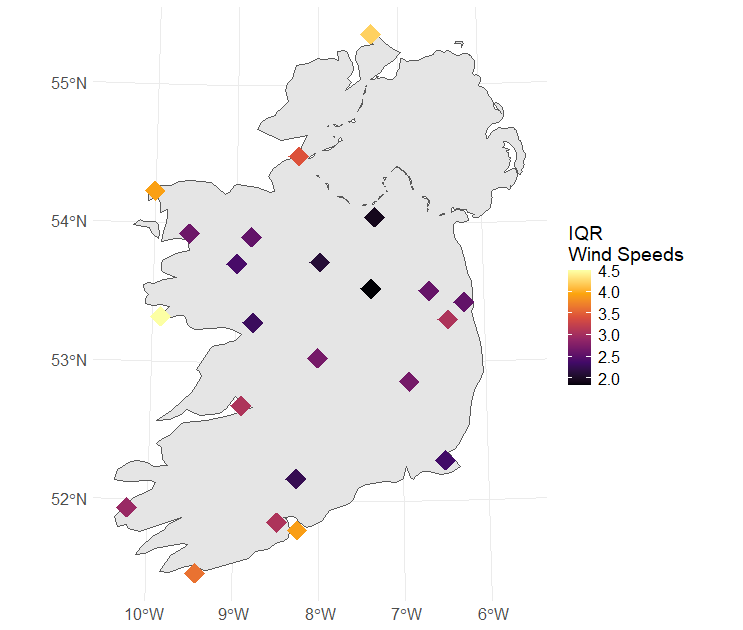}
        \caption{(b)}
        \label{fig:metIQR}
    \end{subfigure}

    \caption{\label{fig:MetStations} The location of Ireland's 23 meteorological stations with: (a) The empirical median at each station and (b) The empirical inter-quartile range~\citep{MetEireannObservation}. The four stations used in Figure \ref{fig:DiurnalWind} are numbered in the green boxes.}
\end{figure}

As Ireland's national meteorological service, Met Éireann maintains a network of quality-controlled weather stations, adhering to the standards set by the World Meteorological Organization~\citep{oke2018guide}. Meteorological station data sourced from Met Éireann is regarded as ground truth in this study due to the high-quality measurement devices used and the optimal siting of instruments, ensuring no obstacles in the vicinity of measurement stations~\citep{MetEireannObservation}. Wind speed measurements are taken at a standard height of 10\,m in open locations, ensuring no obstructions within a 100\,m radius~\citep{MetEireannWind}. The data is available in one minute intervals for 18 stations, and at hourly level for a further 5 stations. For consistency across all stations, including the PWS data introduced in Section~\ref{SectionPWS}, the data is aggregated to an hourly level by averaging sub-hourly measurements over the preceding hour. Aggregating high-frequency wind measurements is a standard practice in wind resource studies, as it balances temporal resolution with noise reduction and compatibility with other datasets \citep{torres2005forecast}. Figure \ref{fig:MetStations} shows the location of the stations, and the median and inter-quartile range of wind speeds over a full year. Although there are only 23 weather stations which measure wind speeds in Ireland—a relatively small number—they are distributed across the country, ensuring good geographic coverage. In general, median wind speeds and wind speed variability are highest in coastal locations, particularly the west coast.

\begin{figure}[!htbp]
\centering
\includegraphics[width=0.7\textwidth]{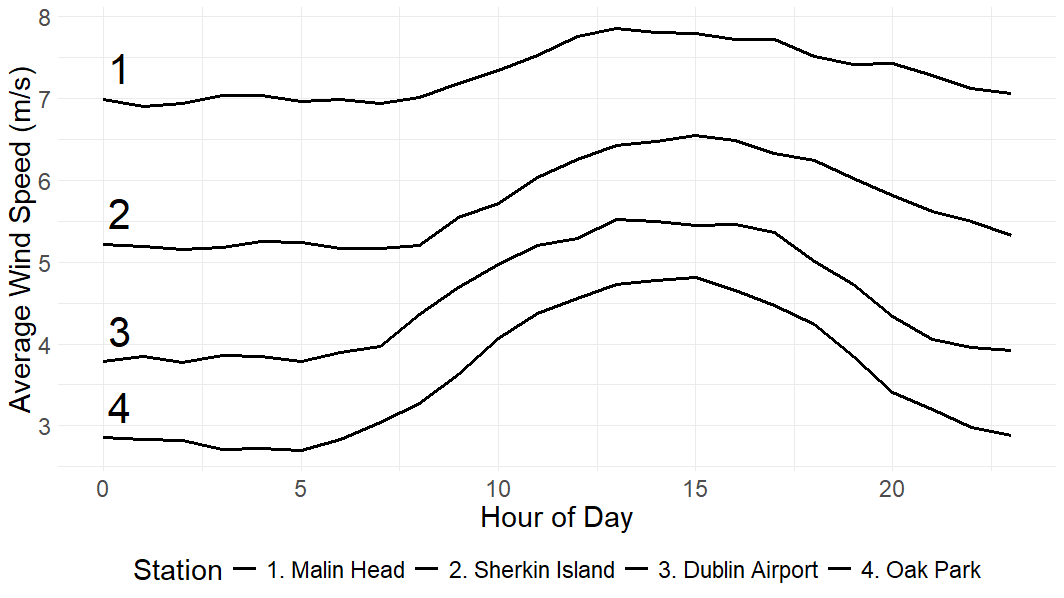}
\caption{\label{fig:DiurnalWind} Hourly wind speed average over the year 2024 at four meteorological stations. The four sample stations are numbered on the map of median wind speeds in Figure \ref{fig:MetStations}.}
\end{figure}

Wind speeds also exhibit a diurnal pattern, as illustrated in Figure \ref{fig:DiurnalWind}, which presents  the average hourly wind speeds over the year 2024 at four station locations. This shows higher wind speeds during the afternoon with lower wind speeds at night, and a stronger diurnal cycle at inland and leeward locations. This pattern is commonly seen in wind speeds and attributed to the effects of solar heating \citep{dai1999diurnal}. 

\FloatBarrier
Wind observations from Irish meteorological stations have been used in previous statistical research studies, and these have demonstrated the strong spatial patterns present in the data~\citep[e.g.][]{haslett1989space,de2005predictive}.

\FloatBarrier
\subsection{Personal Weather Stations}
\label{SectionPWS}
The personal weather station (PWS) data is obtained from \citet{WowMap}, which provides measurements from stations set up by individual users. We selected this dataset because it is hosted on a single platform alongside meteorological station data and includes supporting information on station setup. Other sources of crowdsourced weather data include commercial platforms such as \citet{wunderground} and \citet{Netatmo}, which could be considered in future research. The data from each station is sent digitally to the website which displays weather data and metadata on all stations. Unlike meteorological stations, there is no consistency of equipment or setup, nor is there any quality control. The website only provides information on the location in addition to a broad classification of the site attributes. These fall under three broad classifications, which are based on grading schemes from meteorological bodies \citep{WowGrading},  along with other stations which have not been classified. The three main classifications are described in Table \ref{tab:station_class}:
\begin{table}[h]
\centering
\begin{tabular}{ |c|p{12cm}| }  
 \hline
   Station Class & Description \\
 \hline
 A &  Wind sensors calibrated within the last 10 years, mounted 10m above the ground on mast or pole, with no obstructions within 100m.  \\ 
 B & Wind sensors mounted above the ground on mast or pole, with no obstructions within 50m. \\ 
 C & Wind sensors mounted on building or wall. \\
 U & Wind sensors with unknown set-up attributes \\
 \hline
\end{tabular}
\caption{Classification of PWS by sensor set-up and site attributes}
\label{tab:station_class}
\end{table}

Temporal resolution of the stations can vary, although they are generally at most hourly, and frequently contain data on a sub-hourly resolution, many of which are on a 5-minute resolution. 

\begin{figure}[!htbp]
\centering
\includegraphics[width=0.9\textwidth]{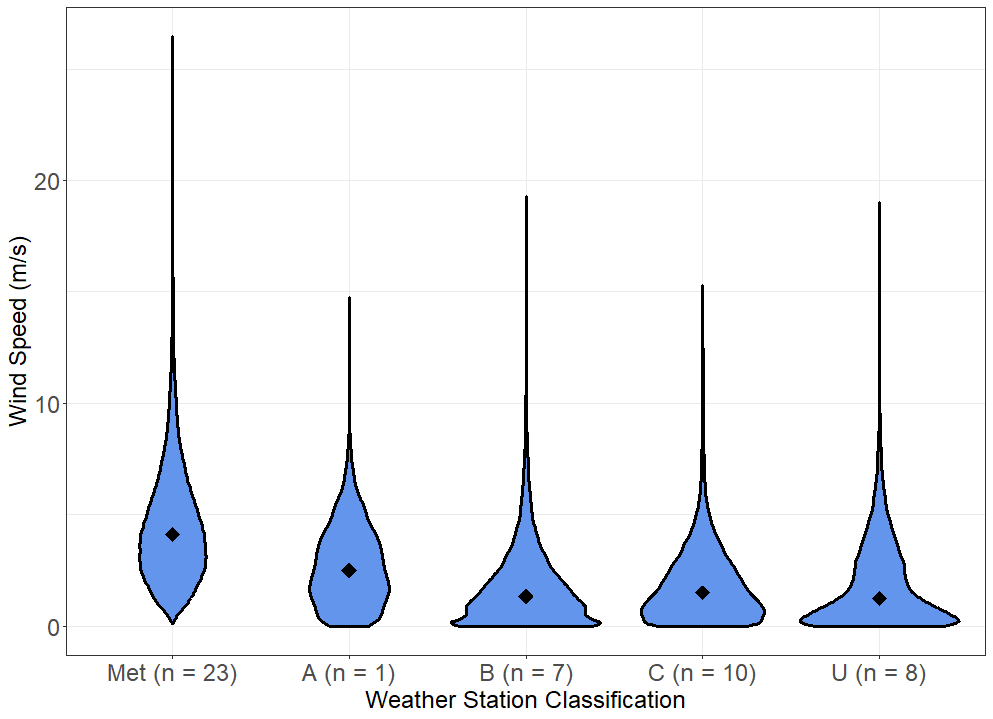}
\caption{\label{fig:Violin} Violin plot showing the wind speed distribution for each group of stations in the Republic of Ireland. A, B, C, and U are the different classes of PWS, Met refers to meteorological weather stations. The black diamonds denote the median of each group. $n$ represents the number of stations in each class.}
\end{figure}

Figure \ref{fig:Violin} shows the distribution of wind speeds across different measurement sources, and highlights that wind speeds at PWS are generally lower when compared to meteorological stations, a similar finding to previous studies on crowdsourced data \citep[e.g.][]{chen2021quality}. This may be due to the high dependency of wind measurements on the placement of equipment. Unlike weather variables such as temperature or air-pressure, local obstacles such as buildings, trees or hedges can have a significant effect on local wind speeds. Meteorological stations typically place anemometers at 10m heights, whereas the heights of the PWS measurements are assumed to be at less than 10m unless the station is in class A.  

\begin{figure}[!htbp]
\centering
\includegraphics[width=0.7\textwidth]{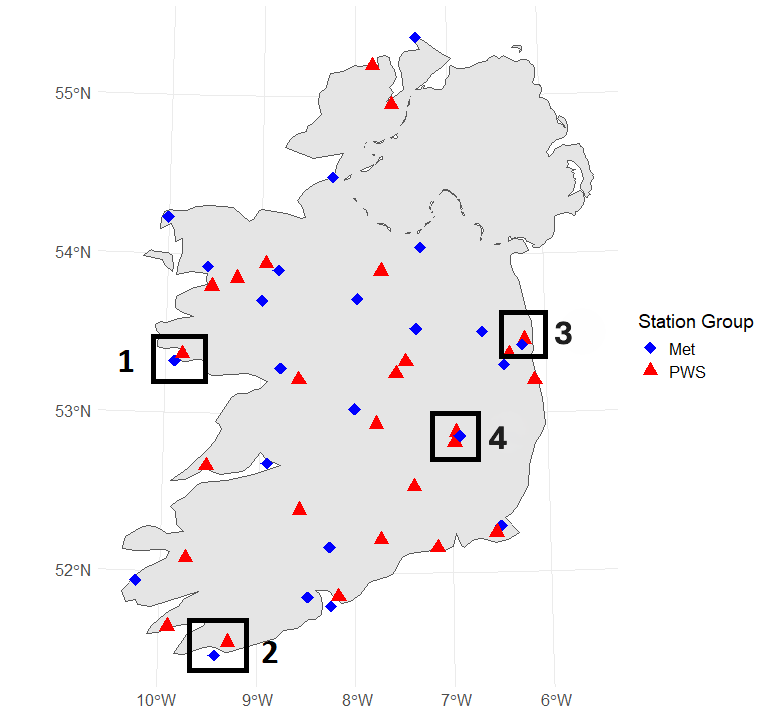}
 \caption{\label{fig:MetPWSLoc} Location of meteorological stations in blue, and PWS in red. Four groups have been highlighted and sample time series for each group are shown in Figure~\ref{fig:OfficialMetRaw}.}
\end{figure}

Consequently, directly including PWS wind speeds in a model, be it statistical or a physical weather model, would most likely lead to significant inaccuracies, in particular underestimation of wind speeds~\citep{chen2021quality}. To highlight the potential usefulness of crowdsourced stations we select examples where we have a meteorological station and PWS in close proximity($<$15km) to each other, presented in Figure \ref{fig:MetPWSLoc}. %In Figure \ref{fig:MetPWSLoc} four groups of neighbouring stations are highlighted as an illustrative example. 

\begin{figure}[H] 
    \centering
    % First row
    \begin{subfigure}[b]{0.45\textwidth}
        \includegraphics[width=\textwidth]{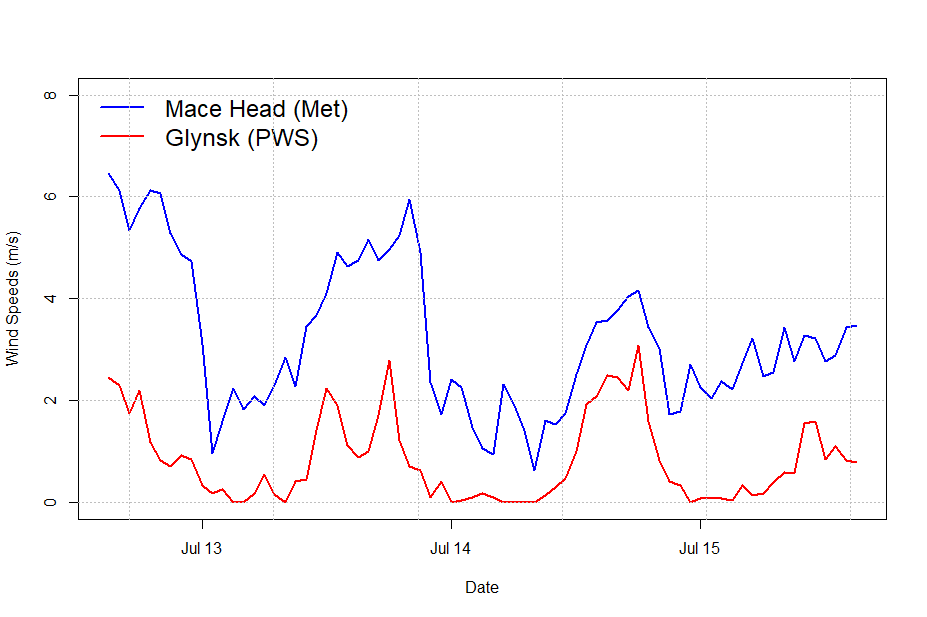}
        \caption{\textbf{1.} Mace Head (Met) and Glynsk (PWS, Class U). Pearson correlation = 0.86}
        \label{fig:MaceGlynskRaw}
    \end{subfigure}
    \hfill
    \begin{subfigure}[b]{0.45\textwidth}
        \includegraphics[width=\textwidth]{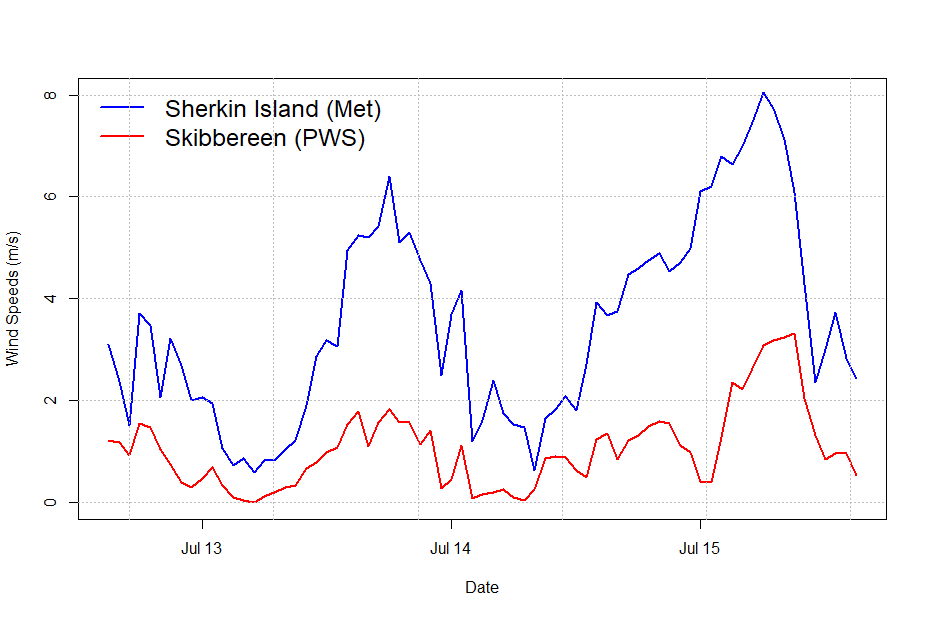}
        \caption{\textbf{2.} Sherkin Island (Met) and Skibbereen (PWS, Class B). Pearson correlation = 0.76}
        \label{fig:SherkinSkibbRaw}
    \end{subfigure}
    
    % Second row
    \vskip\baselineskip
    \begin{subfigure}[b]{0.45\textwidth}
        \includegraphics[width=\textwidth]{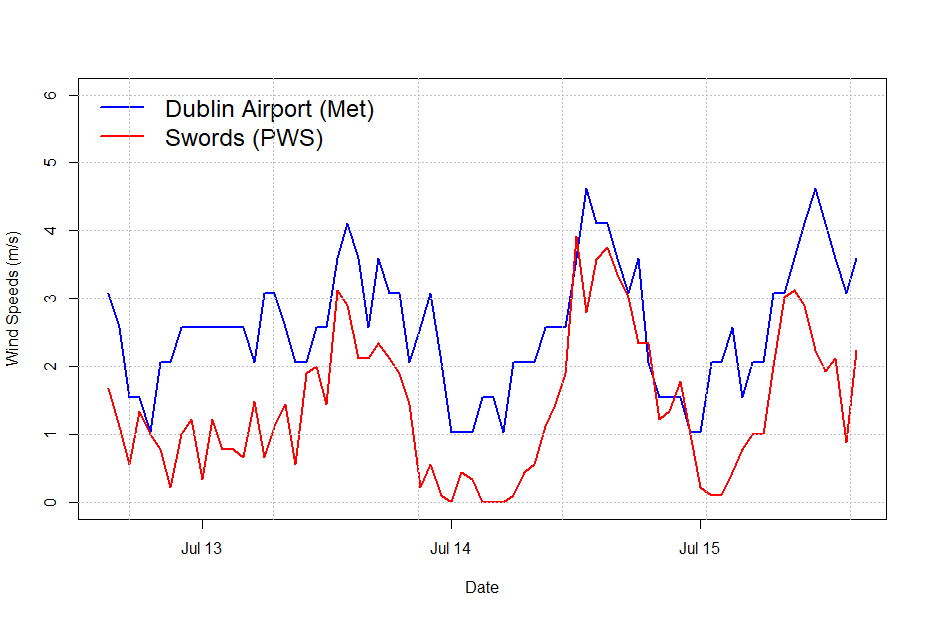}
        \caption{\textbf{3.} Dublin Airport (Met) and Swords (PWS, Class C). Pearson correlation = 0.79}
        \label{fig:DublinSwordsRaw}
    \end{subfigure}
    \hfill
    \begin{subfigure}[b]{0.45\textwidth}
        \includegraphics[width=\textwidth]{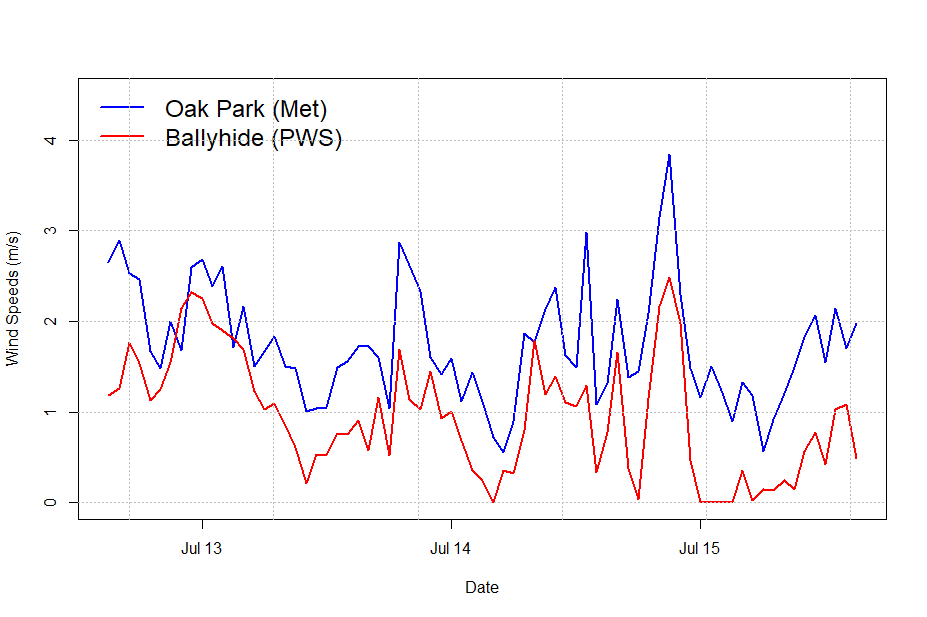}
        \caption{\textbf{4.} Oak Park (Met), Ballyhide (PWS, Class U). Pearson correlation = 0.83}
        \label{fig:Oak2Raw}
    \end{subfigure}
    
    \caption{Comparison of wind speed time series recorded at nearby meteorological (Met) stations (in blue) and PWS (in red) at four locations around Ireland over the period July 13th - July 16th 2024. Panel captions indicate the Pearson correlation coefficient between the two time series.}
    \label{fig:OfficialMetRaw}
\end{figure}

Figure \ref{fig:OfficialMetRaw} compares the wind speeds for our four sets of stations over a three day period. It is clear that the meteorological weather stations have consistently higher wind speeds.

\begin{figure}[H] 
    \centering
    % First row
    \begin{subfigure}[b]{0.45\textwidth}
        \includegraphics[width=\textwidth]{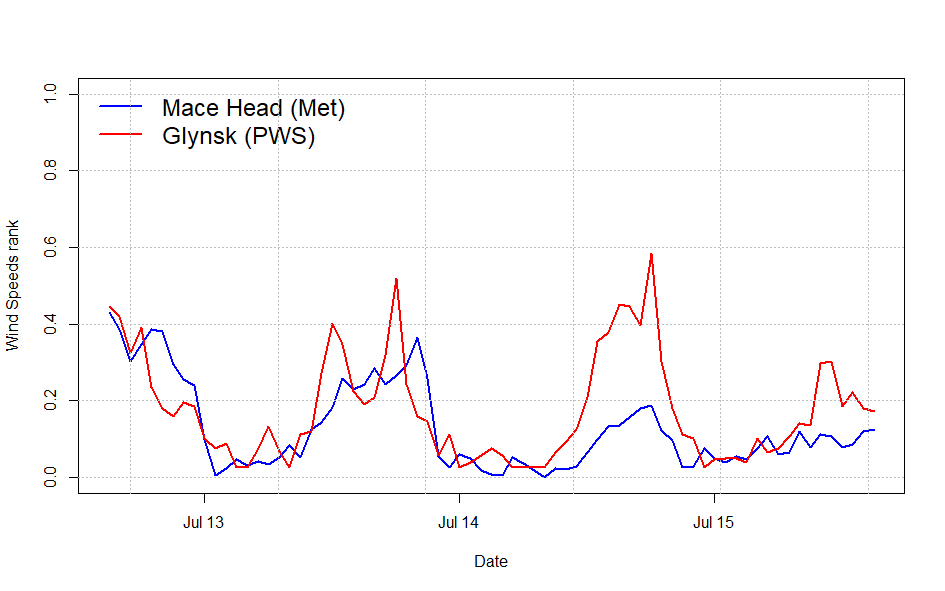}
        \caption{\textbf{1.}  Mace Head (Met) and Glynsk (PWS, Class U). Pearson correlation = 0.85}
        \label{fig:plot1}
    \end{subfigure}
    \hfill
    \begin{subfigure}[b]{0.45\textwidth}
        \includegraphics[width=\textwidth]{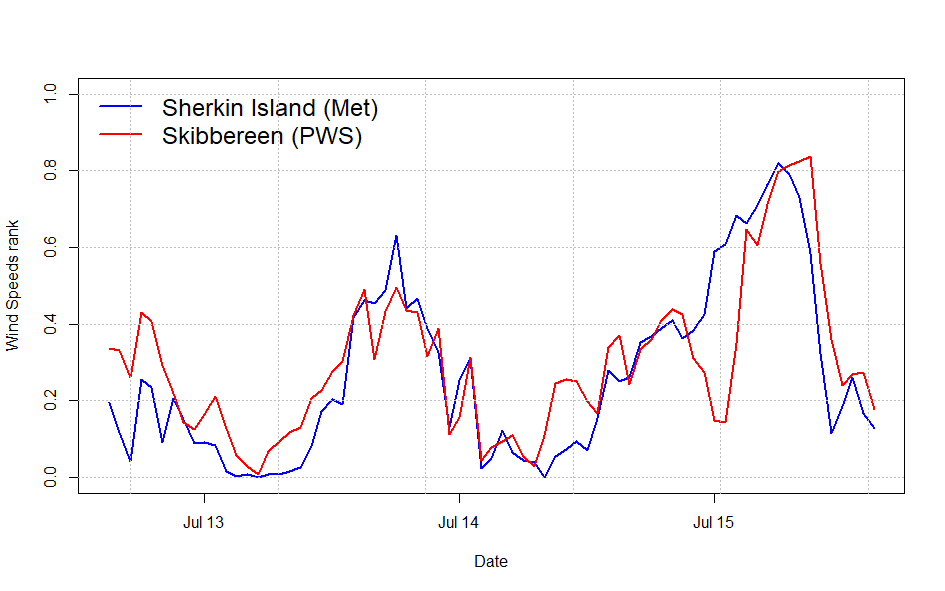}
        \caption{\textbf{2.} Sherkin Island (Met) and Skibbereen (PWS, Class B). Pearson correlation = 0.78}
        \label{fig:plot2}
    \end{subfigure}
    
    % Second row
    \vskip\baselineskip
    \begin{subfigure}[b]{0.45\textwidth}
        \includegraphics[width=\textwidth]{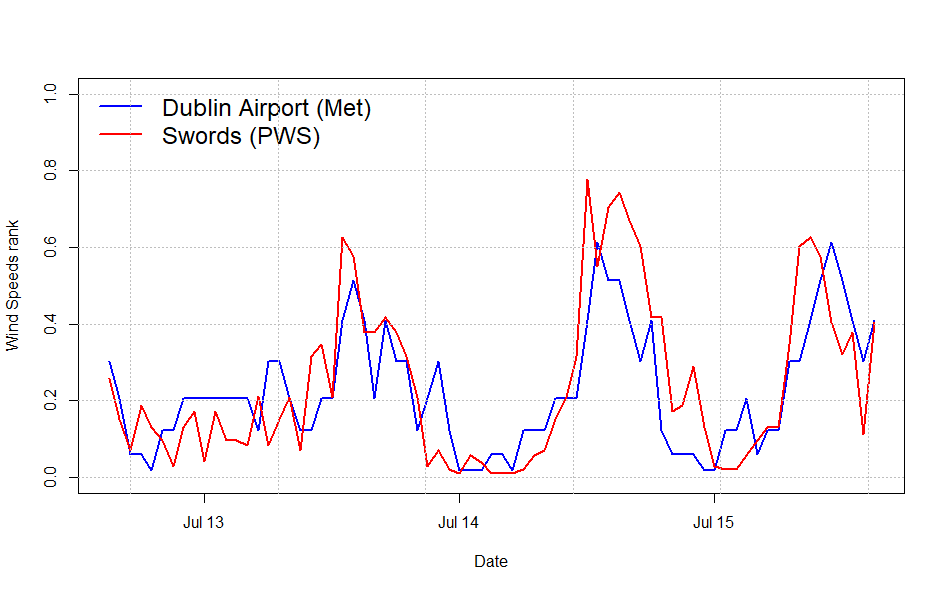}
        \caption{\textbf{3.} Dublin Airport (Met) and Swords (PWS, Class C). Pearson correlation = 0.80}
        \label{fig:plot3}
    \end{subfigure}
    \hfill
    \begin{subfigure}[b]{0.45\textwidth}
        \includegraphics[width=\textwidth]{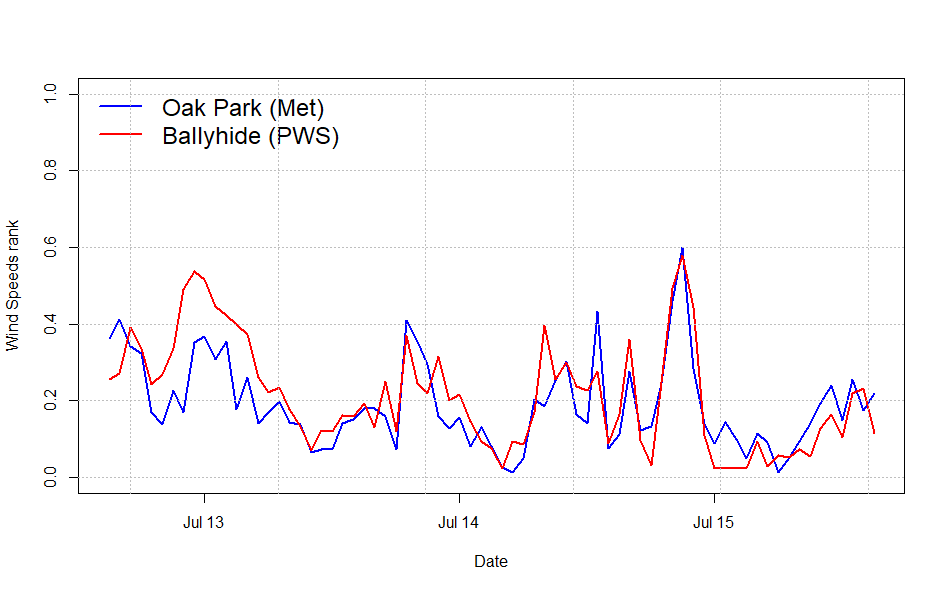}
        \caption{\textbf{4.} Oak Park (Met) and Ballyhide (PWS, Class U). Pearson correlation = 0.83}
        \label{fig:plot4}
    \end{subfigure}
    
    \caption{Comparison of wind speed rank time series recorded at nearby meteorological (Met) stations (in blue) and PWS (in red) at four locations around Ireland over the period July 13th - July 16th 2024. Note that the ECDFs used to compute the ranks cover the entire dataset while the plots display a three-day sample period. Panel captions indicate the Pearson correlation coefficient between the two rank time series, which is equivalent to the Spearman's correlation coefficient between the raw wind speed time series.}
    \label{fig:OfficialPWSQuantiles}
\end{figure}

Although Figure \ref{fig:OfficialMetRaw} illustrates that PWS tend to underestimate the true wind speeds, there appears to be a strong correlation between the time series of the neighbouring stations. To compare measurements from nearby stations on a common scale, we compute the empirical cumulative distribution function (ECDF) for each station, and use this ECDF to transform wind speed time series into rank time series. Such time series express each observation as its rank (normalized in [0,1]) within that station’s ECDF. Figure \ref{fig:OfficialPWSQuantiles} shows that after rank transformation neighbouring time series are in good agreement, and exhibit similar temporal behaviour and comparable ranks at each time point.

In general, PWS data appears to be correlated with nearby meteorological stations. However, as PWS can vary widely in reliability and data quality, we apply a series of quality assurance checks to the raw wind speed time series, and identify and remove stations that do not meet these checks. Bias correction methods for addressing remaining systematic differences are described in Section~\ref{SectionBiasCorrect}. 

First, we exclude PWS stations with excessive missing data. Only stations with at least 90\% non-missing wind speed observations are retained, a criteria used in \citet{lenzi2020spatiotemporal}. In addition, stations that do not upload data in near real-time are excluded because applications in renewable energy require timely updates.

\begin{figure}[!htbp]
\centering
\includegraphics[width=0.7\textwidth]{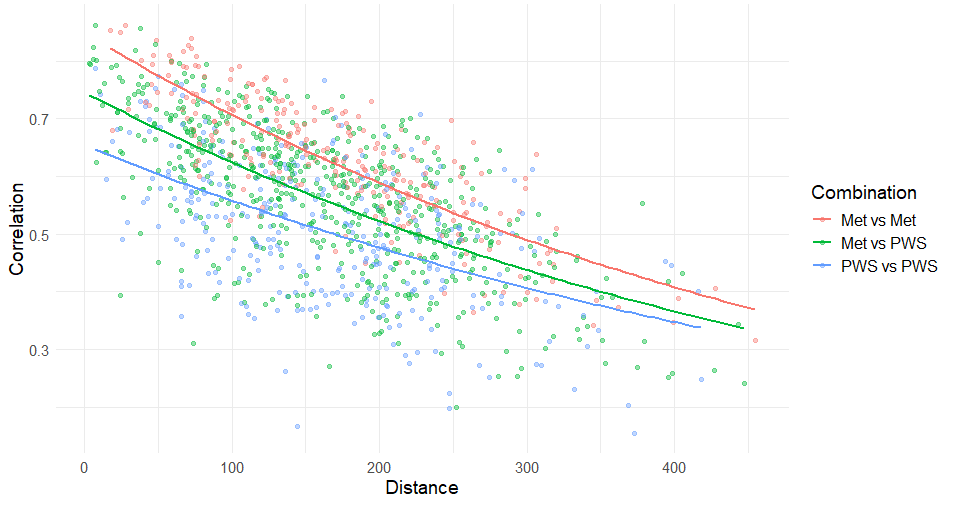}
\caption{\label{fig:SpearmanDistance} Spearman's Rank correlation against distance in kilometres for each pair of stations. Points are coloured based on the classification of the stations. An exponential fit is included for each pairing.}
\end{figure}

Second, we assess data quality using empirical Spearman's rank correlations. Spearman's correlation has been selected because it is robust to non-linear relationships, and therefore captures consistency in wind speed rankings rather than raw values, which is important when stations exhibit non-linear biases as illustrated in Figures~\ref{fig:OfficialMetRaw} and~\ref{fig:OfficialPWSQuantiles}. Figure~\ref{fig:SpearmanDistance} investigates how Spearman's rank correlation is influenced by the distance between stations and the class of stations. As expected, meteorological stations display the highest pairwise correlations with each other, and the pairwise correlation decreases with the distance between the stations. Similarly, the trend of decreasing correlation with distance is seen for Met-PWS and PWS–PWS station pairs. Finally, one can notice that for a given separation distance the Met-Met stations are the more correlated, followed by Met-PWS and PWS-PWS.

To remove unreliable stations, we extend the approach of \citet{chen2021quality}, who discard stations with correlation below 0.5 relative to the nearest meteorological station. In our case, a PWS is retained only if at least five of its nearest neighbours (which may be meteorological or personal stations) exhibit a Spearman correlation greater than 0.5. This threshold is chosen as all meteorological stations meet this criteria, indicating it is a reasonable benchmark.
While we do not correct or remove point-level anomalies in individual time series, this correlation based filtering removes stations that consistently report conflicting observations with nearby quality-controlled stations.

Our final dataset for spatial modelling in Section \ref{SectionStat} contains the 23 meteorological stations in the Republic of Ireland, along with 26 PWS which have been retained after data quality checks have been applied.

\FloatBarrier
\subsection{Reanalysis Data}
\label{Sec:Reanalysis}
Reanalysis data is created by data assimilation techniques using physical weather models combined with historical observations~\citep{lahoz2014data}. They are run retrospectively using observations as boundary conditions and forcings to generate a more accurate and complete picture of historical weather conditions. The output of reanalysis products is a wide range of weather variables on a spatial grid over a long time period, typically at an hourly resolution. However, these outputs are not produced in real time and are usually released with a delay ranging from several days to a few weeks. Commonly used reanalysis datasets are ERA5 from the European Centre for Medium-Range Weather Forecasts (ECMWF)~\citep{hersbach2020era5}, and MERRA2 from NASA ~\citep{gelaro2017modern}. In previous studies, reanalysis data has proven to be an important tool in renewable resource estimation \citep[e.g.][]{doddy2021reanalysis,olauson2018era5}. One of the shortcoming of these reanalysis products is the relatively coarse spatial resolution. For ERA5, the output variables are on a $0.25^{\circ} \times 0.25^{\circ}$ grid, which is approximately $17\text{km} \times 28\text{km}$ at Ireland's latitude. 
\par
\begin{figure}[h!]
        \centering
        \includegraphics[width=0.7\textwidth]{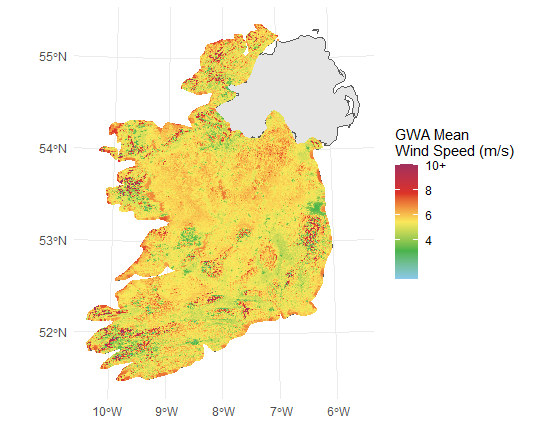}
        \caption{Map of average wind speeds at 10m height on a 250m $\times$ 250m grid, from the Global Wind Atlas \citep{GWADownloads}.}
        \label{fig:GWA}
\end{figure}

In order to better estimate local weather conditions, very high spatial resolution reanalysis products have been created using ERA5 as initial conditions and running weather models at a fine scale. Examples of such datasets include the wind atlas from the Sustainable Energy Authority of Ireland \citep{SEAIWindAtlas}, and the Global Wind Atlas (GWA) \citep{davis2023global}. The SEAI wind atlas is specific to Ireland and designed for wind energy applications, however, it doesn't contain data at 10m, the operational height of weather stations. In this paper, we focus on the GWA due to its global coverage, and it contains data output at 10m heights. Figure \ref{fig:GWA} shows the average 10m wind speed modelled by the GWA. The GWA contains static long-term wind speed statistics, such as average wind speed at several heights above ground level, including 10m heights. The spatial resolution of the GWA is a 250m $\times$ 250m grid. Along with average wind speeds on a fine spatial grid, the GWA also provides the shape and scale parameters for the two parameter Weibull distribution on the 250m grid. The Weibull distribution is a frequently used distribution for modelling wind speeds, and will be discussed further in section \ref{QuantileTransform}. While the spatial resolution is very high, these variables are available only as static, long-term averages; in other words, the GWA does not provide wind speed time series as is the case for the ERA5 and MERRA2 reanalysis datasets, but instead GWA only provides statistics that characterize the probability distribution of wind speed at each grid point and target height. Previous studies, such as \citet{gruber2022towards}, have used the GWA to bias correct coarser reanalysis data to create higher resolution maps. In Section \ref{SectionBiasCorrect} we will use GWA reanalysis data to bias correct the crowdsourced stations, while in section \ref{SectionStat} the GWA data will provide covariates for our models.

\FloatBarrier
\section{Bias Correcting Crowdsourced Weather Data}
\label{SectionBiasCorrect}
Crowdsourced weather data often exhibits bias compared to measurements from meteorological stations, typically showing a downward bias as discussed in Section~\ref{SectionPWS} — a trend also found in previous studies \citep[e.g.][]{droste2020assessing,chen2021quality}. After performing the data quality checks described in Section~\ref{SectionPWS}, we apply a bias correction step before including crowdsourced wind speeds in the spatio-temporal models presented in Section~\ref{SectionStat}.

An alternative approach we considered was scaling the measured wind speeds to match a theoretical mean estimated from reanalysis data (such as the datasets discussed in Section~\ref{Sec:Reanalysis}). However, this method occasionally resulted in unrealistically high corrected values, particularly for stations with very low empirical means. We therefore adopt an alternative method: quantile transformation, which corrects the full distribution of wind speeds and constrains corrected values to remain physically plausible.

\subsection{Selecting optimal distribution}
\label{sec:optimaldist}
To perform the quantile transformation, we examine several parametric distributions to identify the most appropriate for modelling Irish wind speeds. Commonly used distributions in wind speed modelling include the Weibull, log-normal, Gamma, and Rayleigh \citep[e.g.][]{jung2019wind,chen2021assessing,ouarda2015probability}. We focus on the first three distributions, as the Rayleigh is a special case of the Weibull with shape parameter $k = 2$. Table \ref{tab:distributions} provides the probability density functions and parameters for each distribution.

\begin{table}[h]
\centering
\begin{tabular}{|l|l|l|}
\hline
\textbf{Distribution} & \textbf{Probability Density Function} & \textbf{Parameters} \\
\hline
Weibull & 
$f(x) = \frac{k}{\lambda} \left(\frac{x}{\lambda}\right)^{k-1} e^{-(x/\lambda)^k}$ &
$k > 0$ (shape), $\lambda > 0$ (scale) \\
\hline
Gamma &
$f(x) = \frac{\beta^\alpha}{\Gamma(\alpha)} x^{\alpha-1} e^{-\beta x}$ &
$\alpha > 0$ (shape), $\beta > 0$ (rate) \\
\hline
Log-normal &
$f(x) = \frac{1}{x\sigma\sqrt{2\pi}} \exp\left(-\frac{(\ln x - \mu)^2}{2\sigma^2}\right)$ &
$\mu \in \mathbb{R}$ (log-mean), $\sigma > 0$ (log-sd) \\
\hline
\end{tabular}
\caption{Probability distributions considered for wind speed modelling}
\label{tab:distributions}
\end{table}

\begin{figure}[!h]
\centering
\includegraphics[width=0.7\linewidth]{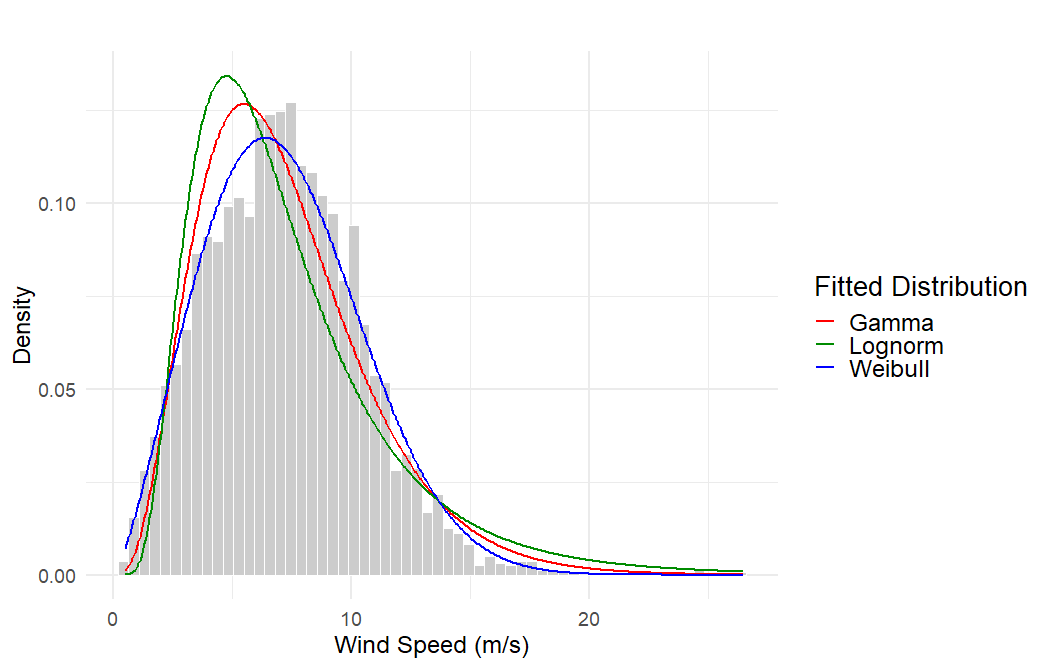}
\caption{\label{fig:MaceDistribution}Three potential distributions along with histogram of wind speeds at Mace head (Site 1 in Figure \ref{fig:MetPWSLoc}).}
\end{figure}

Following previous studies \citep{jung2019wind,chen2021assessing,ouarda2015probability}, we first compare maximised log-likelihoods for each fitted distribution at each station. As each distribution contains two parameters, a penalised criteria such as AIC is not required. The number of stations where each distribution has the highest log-likelihood is recorded. Second, we use the Kolmogorov-Smirnov test \citep{berger2014kolmogorov} to measure maximum distance between empirical and theoretical cumulative distribution functions. Finally, we assess tail behaviour by evaluating the absolute difference between the empirical and theoretical 95\textsuperscript{th} percentiles.
Parameters for each distribution are estimated using a maximum likelihood approach \citep{rockette1974maximum}, implemented using the R package \textit{fitdistrplus} \citep{delignette2015fitdistrplus}. Figure~\ref{fig:MaceDistribution} illustrates the three fitted distributions at one representative site.

\begin{table}[ht]
\centering
\begin{tabular}{|c|l|c|c|c|}  
 \hline
 \textbf{Test} & \textbf{Weibull} & \textbf{Gamma} & \textbf{Log-normal} \\
 \hline
 Highest log-likelihood (count) & 11 & \textbf{12} & 0 \\
 Kolmogorov-Smirnov distance (mean) & \textbf{0.031} & 0.044 & 0.076 \\  
 95\textsuperscript{th} percentile absolute difference (m/s) (mean) & \textbf{0.111} & 0.418 & 1.412 \\ 
 \hline
\end{tabular}
\caption{Goodness of fit for candidate distributions (mean values across 23 stations). For log-likelihood, higher values indicate better performance; for distance and difference metrics, lower values indicate better performance.}
\label{tab:bestdistribution}
\end{table} 

Table \ref{tab:bestdistribution} shows performance for the three distributions, summarised across 23 meteorological stations. From this we see that, for the log-likelihood, the Weibull and Gamma distributions are highest at a similar number of stations. Furthermore, the Weibull distribution outperforms the other two distributions for average KS statistic and also provides the most accurate representation of the 95\textsuperscript{th} percentile. Overall this demonstrates superior overall fit and tail behaviour for the Weibull.

In addition, the Weibull distribution is widely used in wind resource assessment and is directly available in reanalysis products such as the Global Wind Atlas, making it practical for bias correction. This is consistent with previous studies advocating the Weibull distribution for wind speed modelling \citep{carta2009review,wais2017review,bowden1983weibull,ouarda2015probability}.

To further illustrate the suitability of the Weibull distribution, Appendix~\ref{app:WeibullStudy_qq} presents QQ-plots for four representative meteorological stations, showing generally good agreement with the empirical distribution except for several positive outliers.

\subsection{Quantile transformation to the Weibull distribution}
\label{QuantileTransform}
To bias correct the PWS data we utilise a quantile transformation method. Quantile transformations are also commonly used in climate science to bias correct climate models~\citep[e.g.][]{vrac2017influence}. 
Let us consider $\textbf{w}$, a sample of wind speeds of size $n$ collected at a station, $(w_i)_{i=1,n}$ with empirical CDF $F_n(w) = (\sharp (w_i)_{i=1,n} \leq w)/n$, for $w \in [0,+\infty)$. The percentile associated to $w_i$ is $p_i = F_n(w_i)$, and it is given by the rank of $w_i$ in the sample divided by $n$. Hence $1/n \leq p_i \leq 1$.
For our crowdsourced weather data, we calculate empirical CDF of wind speeds separately at each PWS. The percentile value of a datapoint at site $s$ and time $t$ is denoted as $p_{s,t}$. We then assume a theoretical distribution exists for the true wind speed at $s$. As wind speeds are positive, we assume the support of the distribution is $w \in [0,+\infty)$. The CDF of the theoretical true wind speeds is defined as $\hat{F}_{s}\left(w;\Theta\right)$, where $\Theta$ is the vector of parameters characterizing $\hat{F}_s$. Given a recorded wind speed $w_{s,t}$ with associated percentile $p_{s,t}$, the bias-corrected wind speeds can be calculated by performing an inverse CDF transform. Letting $\widetilde{W}_{s,t}$ be the bias corrected wind speeds:
\begin{equation}
    \widetilde{W}_{s,t} = \hat{F}_{s}^{-1}\left(p_{s,t}\right) = \hat{F}_{s}^{-1}\left( F_n(w_{s,t}) \right).
\end{equation}
A parametric distribution is required to perform the quantile transformation, as well as an estimate of $\Theta$, the parameter set, at the location of each PWS. As found in Section \ref{sec:optimaldist}, the Weibull distribution most closely models the measured wind speeds at Irish meteorological stations. Therefore, $\hat{f_{s}}\left(w;\Theta\right)$, which bias corrects the PWS is the Weibull distribution. 

The PDF of a Weibull random variable $X$, with shape parameter $k$, and scale parameter $\lambda$, is given as:
\begin{equation}
    f(x) = \frac{k}{\lambda}(\frac{x}{\lambda})^{k-1} \text{exp}[-(x/\lambda)^k] , x>0, k>0, \lambda>0.
\end{equation}
with CDF:
\begin{equation}
    F(x) = 1-\exp\{-(x/\lambda)^{k}\}
\label{eq:WeibullCDF}
\end{equation}
The expected value of a Weibull distribution in terms of its parameters is given as:
\begin{equation}
\label{eq:WeibullMean}
    \mathbb{E}(X) = \lambda \Gamma\left(1 + \frac{1}{k} \right),
\end{equation}
where $\Gamma\left(\cdot\right)$ denotes the Gamma function, $\Gamma(x) = \int_0^\infty t^{x-1} e^{-t} \, dt$.
%defined as:
%\begin{equation}
%\label{eq:GammaFunction}
%\Gamma(x) = \int_0^\infty t^{x-1} %e^{-t} \, dt
%\end{equation}
In order to carry out the quantile transformation bias correction on wind speeds, the parameters $\lambda$ and $k$ of the Weibull PDF need to be estimated at the location of each PWS. The estimated parameters are used to transform empirical percentiles to corrected wind speeds using an inverse CDF transform. For a Weibull distribution, the inverse CDF of a probability $p \in [0,1)$ defines a quantile given by inverting Equation \eqref{eq:WeibullCDF}:
\begin{equation}
    z = \lambda\left[-\ln(1 - p)\right]^{\frac{1}{k}}
\end{equation}

\FloatBarrier
\subsection{Estimating Weibull Parameters at PWS Locations}
\label{sec:ParameterEstimate}

\subsubsection*{Selecting a Suitable Reanalysis Dataset}
To bias correct the PWS, we require Weibull shape ($k$) and scale ($\lambda$) parameters at each station location. Reanalysis products provide complete spatial coverage over the study area, enabling parameter estimation at PWS sites. We compare two reanalysis datasets: ERA5, which provides historical wind speed time series and was previously shown to be the most accurate for Ireland \citep{doddy2021reanalysis}, and the Global Wind Atlas (GWA), which offers higher spatial resolution (250m) and directly outputs Weibull parameters. 

For each dataset, Weibull parameters are obtained at all meteorological station locations. For the GWA, published Weibull parameters are directly downloaded. These are interpolated to the station locations using inverse-distance weighting. For ERA5, one year of hourly wind speeds preceding the study period is downloaded. Weibull parameters are estimated at each ERA5 grid cell using maximum likelihood and then interpolated to the station locations. Table~\ref{tab:gwa_era5_comparison} reports the correlation between the reanalysis-derived parameters and the station-derived parameters.

\begin{table}[h!]
\centering
\begin{tabular}{|c|c|c|}
\hline
Metric & GWA vs. Stations & ERA5 vs. Stations \\
\hline
Shape parameter correlation ($k$) & \textbf{0.46} & 0.23 \\
Scale parameter correlation ($\lambda$) & \textbf{0.91} & 0.78 \\
\hline
\end{tabular}
\caption{Correlation between reanalysis Weibull parameters and those derived from meteorological station observations.}
\label{tab:gwa_era5_comparison}
\end{table}

The GWA has substantially stronger agreement with meteorological stations for the scale parameter and moderately better correlation for the shape parameter. Although both datasets show limited skill for estimating the shape parameter, due to lower spatial variability, the GWA’s higher resolution provides a closer match overall. For this reason, the GWA is selected as the distribution for bias correction. 
Although high correlation does not imply that the GWA parameters match the station‐derived values in absolute magnitude, it indicates a strong linear association. This, in turn, supports the feasibility of learning a calibration function that maps GWA Weibull parameters to those estimated at meteorological stations.

The full overview of the bias correction approach is displayed in Figure \ref{fig:BiasCorrectionDAG}:
\begin{figure}[h!]
        \centering
        \includegraphics[width=1\textwidth,trim=2cm 4cm 2cm 1cm, clip]{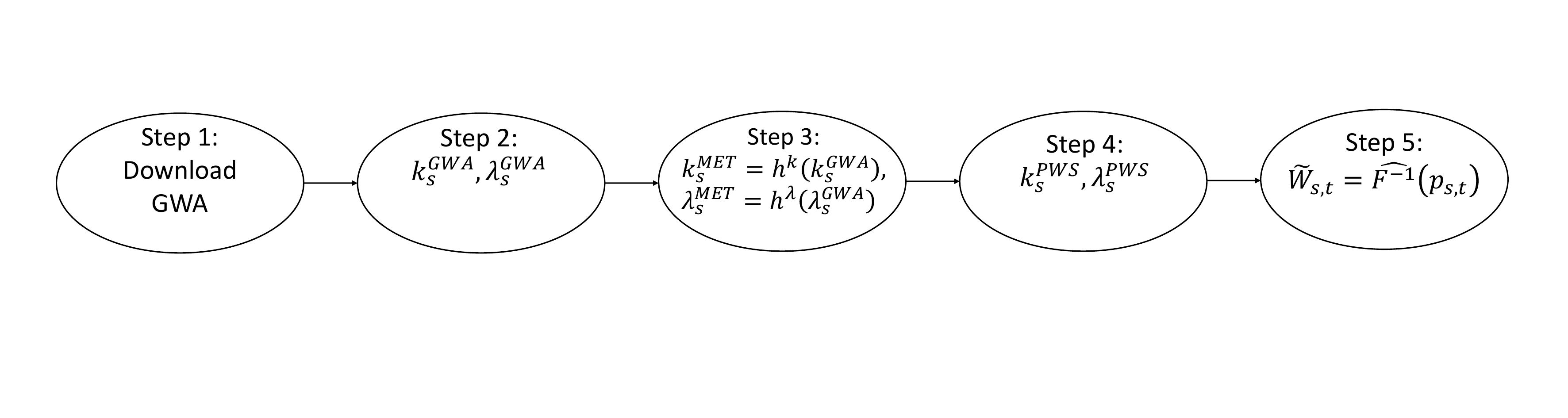}
        \caption{Overview of the five-step bias correction procedure applied to crowdsourced PWS wind speeds. Each step is expanded upon in the following sections.}

        \label{fig:BiasCorrectionDAG}
\end{figure}

\FloatBarrier

\subsubsection*{Step 1. Downloading GWA Data:} The GWA shape ($k$) and scale ($\lambda$) parameters are first downloaded from the online repository \citep{GWADownloads}. These contain both parameters at 10m height on a 250m $\times$ 250m grid.

\subsubsection*{Step 2. Interpolate to weather stations:}
The downloaded parameters are interpolated to meteorological and PWS stations using an inverse distance weighting. Inverse distance weighting is used as the GWA contains over 500,000 data points over Ireland, which prohibits computationally expensive approaches such as Kriging~\citep{wikle2019spatio}. The output of this step is $\{k^{GWA}_s , \lambda^{GWA}_s\}$, the value of the shape and scale from the GWA at the location of both meteorological stations and PWS.

\subsubsection*{Step 3. Calibrating Reanalysis data to Ground Truth:} 
\begin{figure}[!htbp]
    \centering

    \begin{subfigure}{0.45\textwidth}
        \includegraphics[width=\linewidth]{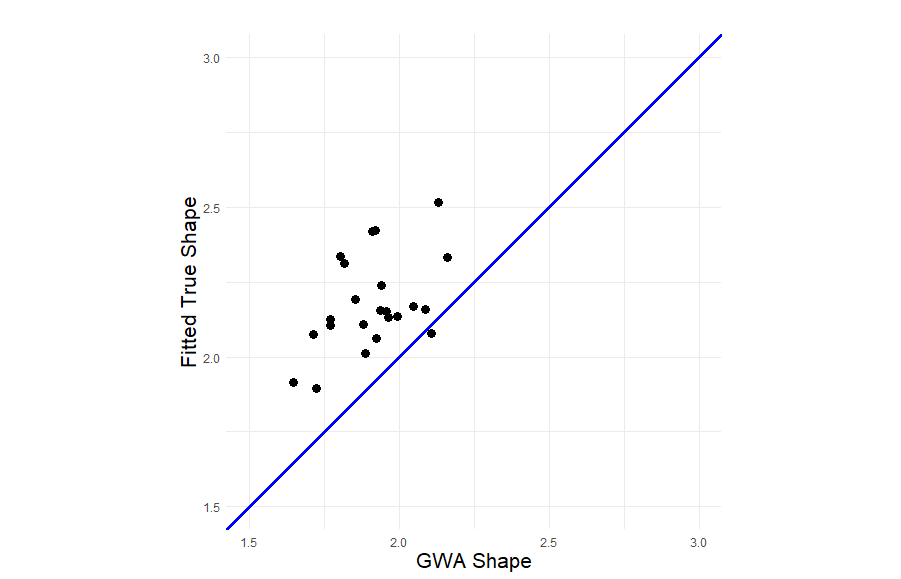}
        \caption{(a)}
        \label{fig:ScaleShapeCompare_a}
    \end{subfigure}
    \hfill
    \begin{subfigure}{0.45\textwidth}
        \includegraphics[width=\linewidth]{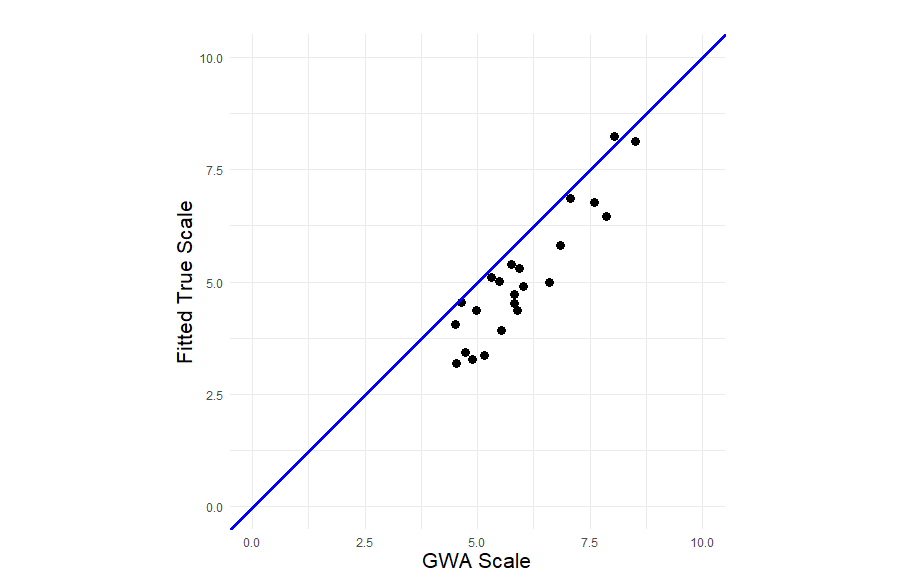}
        \caption{(b)}
        \label{fig:ScaleShapeCompare_b}
    \end{subfigure}

    \caption{Comparison of the parameters of the Weibull distribution derived from the GWA and from observations at the 23 Met Éireann weather stations. (a) Weibull shape parameter. (b) Weibull scale parameter. The blue line in each plot represents the identity line, $y = x$.}
    \label{fig:ScaleShapeCompare}
\end{figure}

To evaluate the accuracy of the parameters from the GWA, we compare them to the Weibull parameters estimated from wind speed time series recorded at meteorological stations, which are considered our ground truth. The Weibull parameters at meteorological stations are estimated using a maximum likelihood approach using the R package \textit{fitdistrplus} \citep{delignette2015fitdistrplus}. Figure \ref{fig:ScaleShapeCompare} compares the shape $\{k^{GWA}_s\}$ and scale  $\{\lambda^{GWA}_s\}$ parameters from GWA to the parameters estimated at the 23 meteorological stations ($\{k^{MET}_s,\lambda^{MET}_s\}$). Results show that the parameters derived from GWA correlate well with the true values but show systematic biases. The GWA underestimates the shape parameter, and overestimates the scale parameter at low-wind sites while remaining accurate at high-wind locations.

\FloatBarrier

\begin{figure}[!htbp]
    \centering
    \begin{subfigure}{0.47\textwidth}
        \centering
        \includegraphics[width=\linewidth]{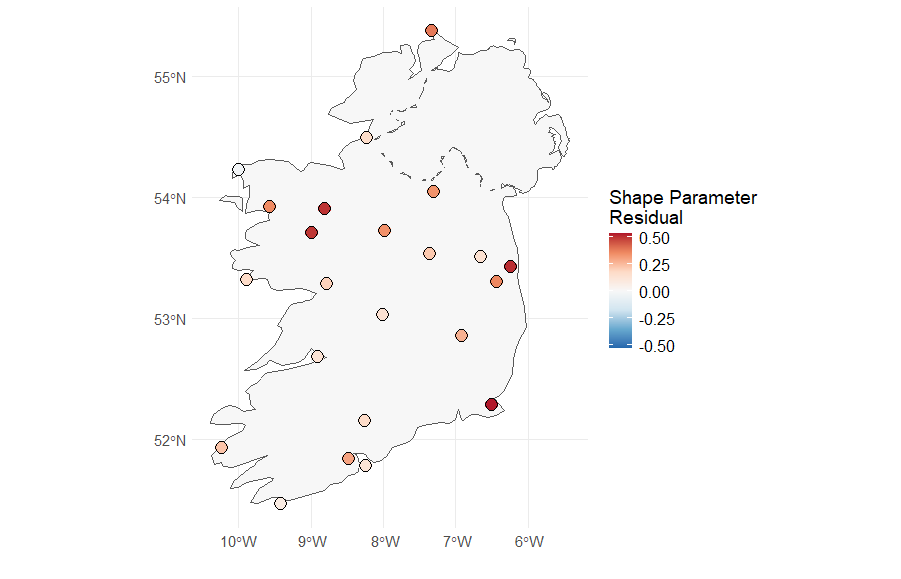}
        \caption{Shape parameter residuals}
        \label{fig:GWA_shape_residuals}
        \end{subfigure}
    \hfill
        \begin{subfigure}{0.47\textwidth}
        \centering
        \includegraphics[width=\linewidth]{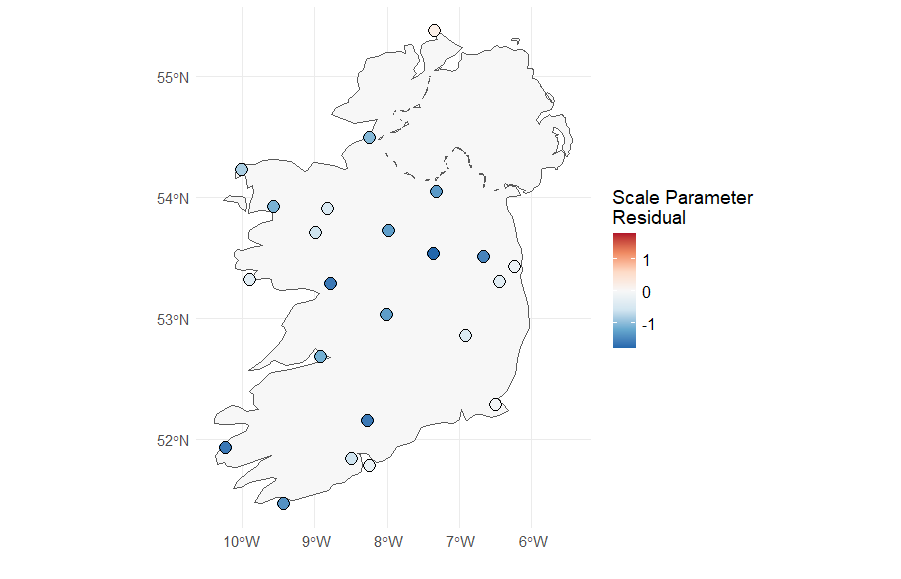}
        \caption{Scale parameter residuals}
        \label{fig:GWA_scale_residuals}
    \end{subfigure}

    \caption{Spatial distribution of the errors (Truth - Prediction) in GWA-derived Weibull parameters. 
    (a) Shape parameter residuals; (b) Scale parameter residuals.}
    \label{fig:GWA_residual_maps}
\end{figure}

To investigate whether the accuracy of the Weibull parameters is spatially dependent, Figure~\ref{fig:GWA_residual_maps} maps the residuals for both the shape and scale parameters. The shape parameter shows no clear spatial structure in its errors. In contrast, the scale parameter exhibits a mild spatial pattern: coastal locations generally display low error, whereas larger errors tend to occur at inland stations. While this pattern is not uniform across all sites, it motivates the inclusion of distance from the coastline as an explanatory variable in the calibration model.

To mitigate the errors of GWA estimates of the Weibull scale and shape parameters, we consider two calibration models, $h^{k}(\cdot)$ and $h^{\lambda}(\cdot)$, which describe the relationship between the GWA parameters at each location and the \textit{true} values derived from meteorological station observations. %Our baseline models were two simple linear regressions with  $\{k^{GWA}_s,\lambda^{GWA}_s\}$ treated as covariates, and $\{k^{MET}_s,\lambda^{MET}_s\}$ as dependent variables. The accuracy of each calibration model was compared using a leave-one out framework, to test how models generalise to unobserved stations. For our shape parameter a linear regression was used as $h^{k}(\cdot)$, the final calibration model: \(k^{MET}_s = \beta_0 + \beta_1k^{GWA}_s + \epsilon^{k}_s\), where \(\epsilon^{k}_s \sim \mathcal{N}(0,\sigma^{2}_{k}) \) is independent Gaussian noise.

According to the structure of errors observed in Figure~\ref{fig:ScaleShapeCompare_a}, and the lack of spatial patterns in Figure~\ref{fig:GWA_shape_residuals}, we selected a simple regression model for the shape parameter, in which $k^{GWA}_s$ is treated as a covariate and $k^{MET}_s$ as the dependent variable. The calibration model for the shape parameter is therefore:
\(h^{k}(\cdot):\hat{k}^{MET}_s = \beta_0 + \beta_1k^{GWA}_s + \epsilon^{k}_s\), where \(\epsilon^{k}_s \sim \mathcal{N}(0,\sigma^{2}_{k}) \) is independent Gaussian noise.

For the scale parameter, the structure of errors observed in Figure~\ref{fig:ScaleShapeCompare_b} and its spatial pattern evidenced in Figure~\ref{fig:GWA_scale_residuals} made us select a generalized additive model (GAM) 
with two covariates: $k^{GWA}_s$ (with thin plate splines as link function) and the distance from the sea (with a linear link). The calibration model for the scale parameter is therefore: \(h^{\lambda}(\cdot):\quad\hat{\lambda}^{MET}_s = \beta_0 + s\left(\lambda^{GWA}_s\right) + \beta_{1}\left(\text{distance from sea (km)}\right)+\epsilon^{\lambda}_s \), where \(\epsilon^{\lambda}_s \sim \mathcal{N}(0,\sigma^{2}_{\lambda}) \) is independent Gaussian noise. 

%For the scale parameter, two extensions beyond linear regression are included. % Firstly to incorporate the spatial dependence seen in Figure \ref{fig:GWAScaleEstimate}, and secondly to include non-linearity in the relationship between $\lambda^{GWA}_s$ and $\lambda^{MET}_s$. 
%To model non-linearity in the relationship between $\lambda^{GWA}_s$ and $\lambda^{MET}_s$, a generalized additive model (GAM) was implemented which used a smooth spline term. %This was fit using the \textit{mgcv} package \citep{wood2015package}. 
%Thin plate splines provide a flexible way of modelling non-linear relationships, while also controlling smoothness to prevent overfitting. To incorporate the dependence seen in Figure \ref{fig:GWAScaleEstimate} distance from the see is added as a linear covariate. The final calibration model, $h^{\lambda}(\cdot)$, for the scale parameter is given as: \(h^{\lambda}(\cdot): \quad \hat{\lambda}^{PWS}_s = \beta_0 + s\left(\lambda^{GWA}_s\right) + \beta_{1}\left(\text{distance from sea (km)}\right)+\epsilon^{\lambda}_s \), where \(\epsilon^{\lambda}_s \sim \mathcal{N}(0,\sigma^{2}_{\lambda}) \) is independent Gaussian noise.

%The full results for the different models are included in Appendix~\ref{BiasCorrection}, along with estimated coefficients.
More details on calibration models selection and estimated coefficients are included in Appendix~\ref{BiasCorrection}. 
To assess the suitability of the calibrated Weibull distribution derived from the GWA for bias correction, we compared the implied distributional properties against empirical wind speed statistics at the meteorological stations. Three aspects of the wind speed distribution were evaluated: (i) the mean, (ii) the variance, and (iii) tail behaviour, assessed using the 95\textsuperscript{th} percentile.

\begin{table}[h!]
\centering
\begin{tabular}{|l|c|c|}
\hline
\textbf{Metric} & \textbf{Mean Absolute Error (MAE)} & \textbf{Pearson Correlation} \\
\hline
Mean Wind Speed & 0.38 & 0.94 \\
Variance & 0.61 & 0.95 \\
95\textsuperscript{th} Percentile & 0.65 & 0.94 \\
\hline
\end{tabular}
\caption{Comparison between empirical wind speed statistics and those implied by the calibrated Weibull distribution across stations. Metrics are aggregated over the 23 meteorological stations.}
\label{tab:weibull_validation}
\end{table}
Table~\ref{tab:weibull_validation} highlights that the calibrated Weibull distributions accurately capture key characteristics of the observed wind speed distributions. Across stations, the implied means, variances, and 95\textsuperscript{th} percentiles show high correlation (0.94–0.95) with empirical values and low mean absolute errors. This indicates that the Weibull representation is consistent with the observed wind speeds and suitable as a target distribution for the bias-correction procedure.

\subsubsection*{Step 4. Predicting Parameters at PWS Locations:} 

The final calibration models are used to predict the shape and scale parameters at PWS locations:

\begin{equation}
\label{eq:ShapeModel}
    h^{k}(\cdot): \quad \hat{k}^{PWS}_s = \beta_0 + \beta_1k^{GWA}_s
\end{equation}

%For the shape parameter, a simple linear regression relates the GWA estimate at each location to the predicted value at the PWS.

\begin{equation}
\label{eq:ScaleModel}
    h^{\lambda}(\cdot): \quad \hat{\lambda}^{PWS}_s = \beta_0 + s\left(\lambda^{GWA}_s\right) + \beta_{1}\left(\text{Distance from sea (km)}\right)
\end{equation}

%For the scale parameter, \( s(\lambda^{GWA}_s) \) is a thin plate spline function of the GWA scale estimate, while the distance from the sea is included as a linear term. These models produce a predicted Weibull distribution for wind speed at each PWS. %The point estimate of the parameters at each location is considered, and in this paper we don't incorporate uncertainty when performing the bias correction.

\subsubsection*{Step 5. Correcting Wind Speeds:}
The corrected wind speeds, $\widetilde{W_{s,t}}$, are obtained using an inverse quantile transform based on their empirical CDF.
\begin{equation}
    \widetilde{W}_{s,t} = {\lambda^{PWS}_s}\left[-\ln(1 - {p_{s,t}}) \right]^{\frac{1}{{k^{PWS}_s}}},
\end{equation}
where $p_{s,t} \in (0,1)$ is the empirical percentile at location $s$ and time $t$. Figure \ref{fig:InverseCDF} in Appendix~\ref{BiasCorrection} shows an example of inverse quantile mapping. Following this step we have completed our bias correction, and now have a dataset of bias corrected PWS.

\begin{figure}[H] 
    \centering
    % First row
    \begin{subfigure}[b]{0.45\textwidth}
        \includegraphics[width=\textwidth]{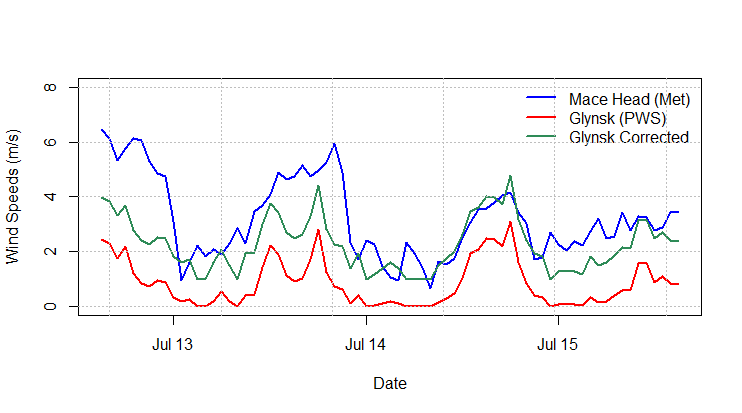}
        \caption{\textbf{1.} Mace Head (Met) in blue, compared to Glynsk (PWS) in red and green.}
        \label{fig:MaceGlynskRawCorrected}
    \end{subfigure}
    \hfill
    \begin{subfigure}[b]{0.45\textwidth}
        \includegraphics[width=\textwidth]{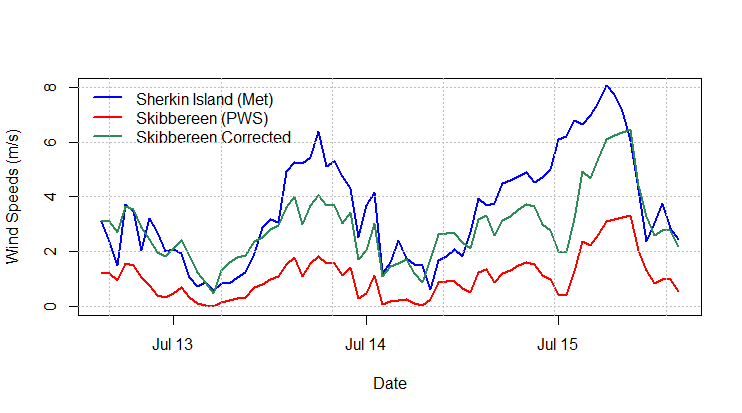}
        \caption{\textbf{2.} Sherkin Island (Met) in blue, compared to Skibbereen (PWS) in red and green.}
        \label{fig:SherkinSkibbRawCorrected}
    \end{subfigure}

    % skip row
    \vskip\baselineskip
    \begin{subfigure}[b]{0.45\textwidth}
        \includegraphics[width=\textwidth]{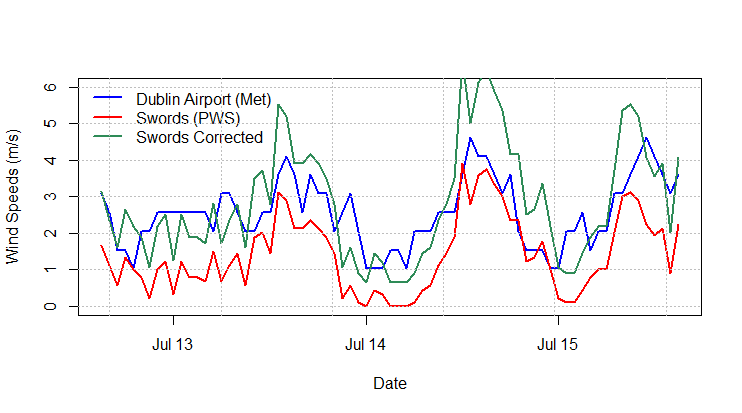}
        \caption{\textbf{3.} Dublin Airport (Met) in blue, compared to Swords (PWS) in red and green.}
        \label{fig:DublinSwordsrawCorrected}
    \end{subfigure}
    \hfill
    \begin{subfigure}[b]{0.45\textwidth}
        \includegraphics[width=\textwidth]{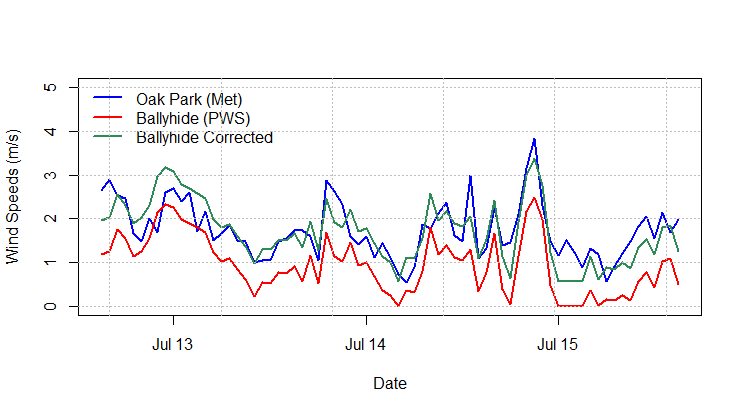}
        \caption{\textbf{4.} Oak Park (Met) in blue, compared to Ballyhide (PWS) in red and green.}
        \label{fig:OakBallyhiderawCorrected}
     \end{subfigure}
    
    \caption{A comparison of nearby meteorological stations (blue) and PWS (red) at four locations in Ireland. The bias corrected wind speed is added to each plot (green line).}
    \label{fig:OfficialMetRawCorrected}
\end{figure}

Since there is no ground truth at PWS, the accuracy of bias correction cannot be quantified directly. However, Section \ref{SectionResults} will demonstrate the benefits of incorporating bias-corrected stations into spatial models. To visualize the impact of the corrections, Figure \ref{fig:OfficialMetRawCorrected} examines the four pairs of stations from Figure \ref{fig:OfficialMetRaw}, where the corrected wind speeds show closer agreement with neighbouring meteorological stations.

\begin{figure}[!htbp]
\centering
\includegraphics[width=0.7\textwidth]{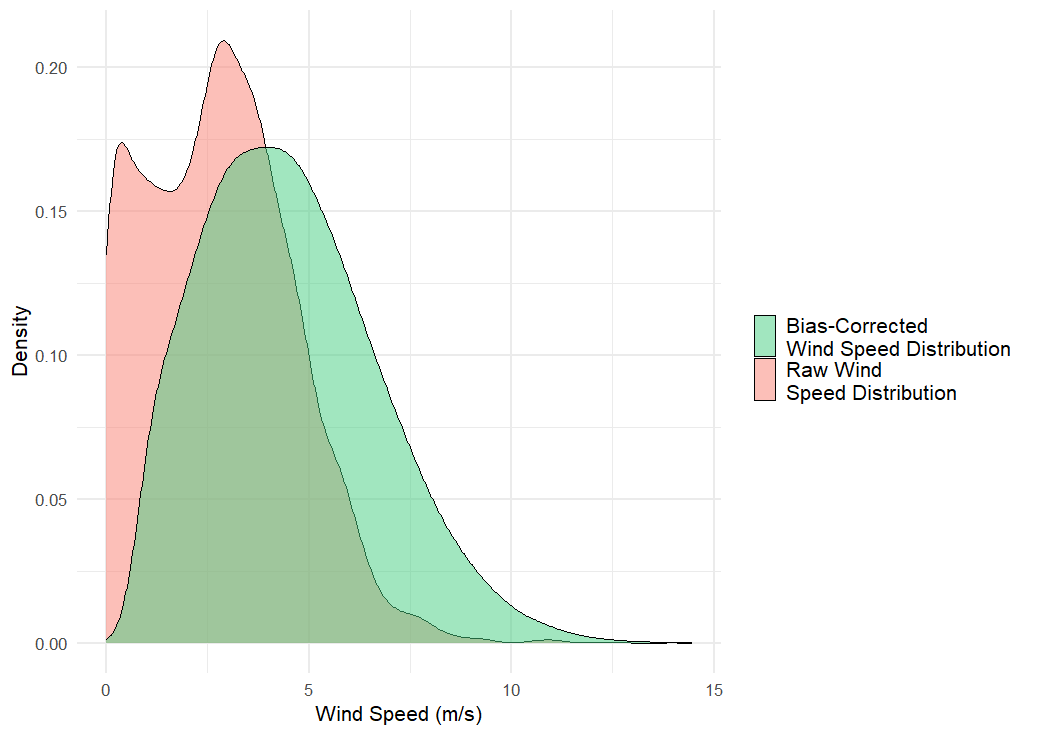}
\caption{\label{fig:CorrectedDist} Sample correction at Glynsk PWS (labelled 1 in Figure \ref{fig:MetPWSLoc}). Kernel density estimates of the raw wind speeds (red), along with bias corrected wind speeds using Weibull quantile transformation (green).}
\end{figure}

To compare the distribution of bias-corrected and raw wind speeds at a sample site, the PWS labelled 1 in Figure \ref{fig:MetPWSLoc} is chosen, and a kernel density estimate of raw wind speeds and corrected wind speeds are shown in Figure \ref{fig:CorrectedDist}. The corrected distribution shifts the wind speed distribution to the right. As expected, the quantile transformation results in wind speeds which closely follows a Weibull distribution. As the quantile transformation maps the empirical percentiles to a parametric distribution, isolated extreme values in the raw data have reduced impact on the overall model. Outlying raw observations are still mapped to high percentiles, but their absolute magnitude becomes consistent with the overall fitted distribution.

Note that while the raw wind speeds appear to have a slight bimodal distribution, this could be due to PWS often storing only discrete values or exhibiting zero inflation. One potential limitation of the quantile transformation is its tendency to underestimate very high wind speeds, particularly those with negligible probabilities under the Weibull distribution. The impact of extremes is further discussed in the Section \ref{sec:extremes}. Based on visual verification, the quantile transformation generally produces wind speeds at PWS that closely match those from neighbouring meteorological stations.

\FloatBarrier
\section{Spatio-Temporal Modelling with Crowdsourced Wind Observations}

Our goal is to investigate the impact of incorporating data from crowdsourced weather stations in the spatial modelling of wind speeds, and to explore modelling strategies that account for potentially noisy data. We focus on predicting wind speeds at unobserved sites using a computationally efficient framework that enables the generation of long-term, hourly wind speed time series at high spatial resolution. The approach also supports real-time mapping of wind speeds across Ireland, making it suitable for both retrospective analysis and operational use.

\label{SectionStat}
\subsection{Spatio-Temporal Models for Wind Speeds}
\label{sec:SectionGP}
Let $W_{s,t}$ be the wind speed at location $s$ and time $t$. This includes our combined datasets of meteorological stations and bias corrected PWS (to simplify notation, the bias corrected wind speeds, $\widetilde{W}_{s,t}$, will also be denoted as $W_{s,t}$). To approximate normality in the right-skewed wind speed data, we apply a square root transformation, a common practice in wind modelling~\citep{lenzi2020spatiotemporal,haslett1989space}. An alternative method would be to transform each station’s observations to have a marginal $\mathcal{N}(0,1)$ distribution using the Weibull CDF discussed in Section~\ref{sec:ParameterEstimate}, but in practice we found little differences in predictive accuracy between the two approaches (see Appendix~\ref{app:N01_appoach} for details), and therefore we decided to adopt the widespread square root transformation. Subsequently the transformed values are decomposed into fixed and random effects.

\begin{equation}
    \sqrt{W_{s,t}} = \mu_{s,t} +d(t)+ z_{s,t} + \epsilon_{s,t}  
\end{equation}
where:
\begin{itemize}
    \item $\mu_{s,t}$ includes the fixed effects describing the process mean
    \item $d(t)$ represents the diurnal temporal trend
    \item $z_{s,t}$ is a zero mean spatio-temporal process
    \item $\epsilon_{s,t}$ represents the uncorrelated, independent noise
\end{itemize}

We further assume $\mu_{s,t}$ is a linear additive combination of $P$ covariates:
\begin{equation}
    \mu_{s,t} = \beta_{0}+ \sum_{j=1}^{P} \beta_j X_{j,s,t},
\end{equation}
where $X_{j,s,t}$ is the $j^{th}$ covariate at location $s$ and time $t$. The covariates can be spatially varying (e.g., elevation), temporally varying (e.g., time of day), or vary in space and time (e.g., temperature). $\beta_j$ is the corresponding coefficient for each covariate. In previous studies on wind energy, a variety of topographical covariates have been used \citep[e.g.][]{lee2022long}, however in our analysis the relationship between these covariates and wind tends to be complex and nonlinear, therefore incorporating them in an additive linear model would fail to model their true effects. Instead for our spatially varying covariates we will use the GWA output which was used for bias correction in Section \ref{SectionBiasCorrect}. Reanalysis datasets are developed using physical models, and these capture how weather variables interact with the topography. However as we are modelling the square root of wind speed, the average wind speed derived from the GWA is not strictly linearly correlated to the square root of wind speeds. To include the GWA output as a covariate, we calculate the mean value of the square root of wind speeds. This is calculated with respect to the Weibull distribution estimated at each spatial location (discussed in Section \ref{QuantileTransform}). The square root of a Weibull distribution is also a Weibull distribution, with $\text{scale}=\sqrt{\lambda}$ and $\text{shape}=2k$, therefore the mean is known in closed form (See Appendix \ref{App:sqrt_Weibull} for derivation).
\begin{equation}
    X_{1,s} = \mathbb{E}\left(\sqrt{W_{s}}\right) = \sqrt{\hat{\lambda}_{s}}\ \Gamma\left(1 + \frac{1}{2\hat{k}_{s}}\right),
\end{equation}
Here $\hat{k}_{s}$ and $\hat{\lambda}_{s}$ are the estimated Weibull parameters at each location.
Previous studies have used other weather variables as covariates, such as temperature or wind direction \citep[e.g.][]{lenzi2020spatiotemporal}, however since the goals is to predict wind speed at an unobserved site, we will assume these are not available. In practice there would be numerical weather predictions of weather variables. However, since these are still based on models and therefore have uncertainty associated with them, we choose not to include them in this study. 

To capture the time-of-day effect, we model a smooth diurnal trend using a cyclic random walk of order 1, following~\citet{gomez2020bayesian}. Let \( t \in \{1, \dots, 24\} \) represent the hour of the day. We include a latent effect \( d(t) \) modelled as:
\begin{equation}
    d(t) - d(t-1) \sim \mathcal{N}(0, \sigma^{2}_{d}) \quad \text{with}\quad d(1)-d(24) \sim \mathcal{N}(0, \sigma^{2}_{d})
\end{equation}
Wind speeds in Ireland exhibit clear seasonal patterns, as well as longer-term variability associated with climate change \citep{leahy2013persistence,nolan2012simulating}. In principle, these effects could be incorporated through a cyclic annual random walk for seasonal variation and an additional long-term temporal component to capture climate-driven trends. However, reliable estimation of these components typically requires several years of data in order to separate long-term trends from short-term fluctuations, whereas our study spans only six months. For this reason, we restrict temporal variation to the diurnal cycle. Seasonal variations within the study period will also be partially accounted for by the spatial random effects, as measured observations will reflect the prevailing weather conditions. 
The final model including covariates is given as:
\begin{equation}
\label{eq:finalmodel}
    \sqrt{W_{st}} = \beta_{0} + \beta_{1}X_{1,s} +d(t) + z_{s,t} + \epsilon_{s,t},
\end{equation}
where $z_{s,t}$ is assumed to be a zero mean Gaussian Process (GP) with covariance function $C$ as given below. \citet{williams2006gaussian} contains a detailed overview of Gaussian Processes. To ensure that the covariance matrix is positive semi-definite, parametric covariance functions are typically used ~\citep{banerjee2003hierarchical}. Within this paper, the Matérn covariance function is used to model the spatial covariance of $z_{s,t}$ and we also assume that the process is stationary and isotropic in space. While this may not be realistic for Irish wind speeds, this assumption allows for a more parsimonious model and can prevent overfitting \citep{fuglstad2015does}. The Matérn covariance function $C(\cdot)$, is defined in \eqref{eq:2}:
\begin{equation}\label{eq:2}
     \text{cov}(z_{i,t},z_{j,t})= C(h_{ij}) = \sigma^2_{z} \frac{2^{1-\nu}}{\Gamma(\nu)} \left( \kappa h_{ij} \right)^\nu K_\nu \left(  \kappa h_{ij} \right),
\end{equation}
where:
\begin{itemize}
    \item $h_{ij}$: Distance between two points in the spatial domain.
    \item $\sigma^2_{z}$: Variance parameter, representing the sill or overall variance of the spatial process.
    \item $\nu$: Smoothness parameter, controlling the differentiability of the spatial field. This is fixed to 1 in our implementation, which is commonly used to balance flexibility and computational efficiency. Additionally in the inference approach discussed in Section \ref{sectionInference}, it is recommended $\alpha = \nu + d/2$ is an integer, where $d=2$ is the number of dimensions \citep{lindgren2011explicit}.
    \item $\kappa$: Scale parameter, controlling the spatial correlation length. Smaller $\kappa$ values indicate stronger correlations over longer distances. In particular the effective range, $\phi$, of the spatial field, is defined as $\phi = \sqrt{8}/\kappa$.
    \item $K_\nu$: Modified Bessel function of the second kind of order $\nu$.
\end{itemize}

As the wind speed observations are indexed by both space and time, we investigate how to model $z_{s,t}$ spatially and temporally. However, since our dataset encompasses over 200,000 datapoints, we seek a computationally efficient and low-memory inference approach. To strike a balance between computational efficiency and model complexity, we consider two approaches. The first assumes temporal independence, treating each time point as an independent replicate of a purely spatial process with covariance function $C(\cdot)$. We call this model the Independent Gaussian Process model (\textbf{IGP}). Although this overlooks temporal correlations, it simplifies computations and enables efficient interpolation across historical time series by dealing with only an $n \times n$ covariance matrix for $n$ spatial locations. 

The second approach is to capture temporal dependencies while maintaining computational feasibility, which is achieved by extending our model to a separable spatio-temporal framework. This assumes the space-time correlation can be factored into a spatial component, modelled with the Matérn covariance function described above, and a temporal component, modelled as an autoregressive process \citep{cameletti2011comparing, lenzi2020spatiotemporal}. We denote this model as \textbf{AR1}.
The temporal correlation between time points $t$ and $t'$ is given by:
\begin{equation} \text{Corr}(t,t') = \rho^{|t-t'|}, \end{equation} where $\rho$ represents the correlation between consecutive time steps. The full space-time covariance is then:
\begin{equation} \text{Corr}((s,t),(s',t')) = C(|s-s'|)\rho^{|t-t'|}. \end{equation} A key advantage of separability is its computational efficiency, as it enables Kronecker product decompositions that simplify covariance matrix operations \citep{wikle2019spatio}. While this assumption may not fully capture complex dependencies, studies such as \citet{genton2007separable} show that it provides substantial computational savings with minimal loss in accuracy for wind speed modelling.

Lastly, $\epsilon_{s,t}$ represents the uncorrelated noise at location $s$ and time $t$, often defined as the nugget effect. It describes the small scale variability not captured by the spatial component, or the measurement error associated with an observation. This is typically assumed to be zero mean and normally distributed:
\begin{equation}
\label{eq:nuggetdef}
    \epsilon_{s,t} \sim \mathcal{N}(0 , \sigma_{\epsilon}^{2}).
\end{equation}
In Equation \eqref{eq:nuggetdef}$, \sigma_{\epsilon}$, which represents the standard deviation of independent noise will be referred to as the nugget variance. It is generally assumed to be constant across all observations. The primary extension in our spatial model is to allow $\sigma_{\epsilon}$ to vary across different observation groups. This is to account for certain observations, such as PWS, having varying levels of measurement error. Allowing $\sigma_{\epsilon}$ to vary can account for potentially noisier sources of data, and reduce their contribution to the prediction when more reliable sites are in close proximity. In this paper we will consider five separate sources of observation, each with a separate independent noise parameter, described in Table \ref{tab:station_noise}
\begin{table}[h]
    \centering
    \begin{tabular}{|l|l|}
        \hline
        \textbf{Category} & \textbf{Description} \\
        \hline
        \textbf{Met} & Observations from meteorological stations (see Section \ref{MetData}). \\
        \textbf{A, B, C} & Separate parameters for three grades of PWS. \\
            & A description of each group is in Table \ref{tab:station_class}. \\
            & Different sensor setups may influence measurement noise. \\
        \textbf{U} & Separate noise parameter for PWS without site grading. \\
        \hline
    \end{tabular}
    \caption{Classification of weather stations and their assigned noise parameters. The station categories are discussed in further detail in Section \ref{SectionPWS}.}
    \label{tab:station_noise}
\end{table}

To define our model for separate noise parameter, we denote \textbf{W} as the set of all observations, and it is partitioned into five groups.
\begin{equation}
    \mathbf{W} = \left[W_{met} , W_{A} , W_{B} , W_{C},W_{U}  \right].
\end{equation}
Each observation can be written in terms of fixed and random components, where cat denotes the station category described in Table \ref{tab:station_noise}:

\begin{align}
    \sqrt{W_{s,t}^{cat}} = \mu_{s,t} +d(t)+z_{s,t} + \epsilon_{s,t}^{cat}\\
\end{align}
Each $\epsilon_{s,t}$,  is assumed to be an independent zero-mean Gaussian distribution (Equation \eqref{eq:nuggetdef}), where the set of $\sigma_{\epsilon}$, the standard distribution, contains:
\begin{equation}
\label{eq:sigmaparams}
    \sigma_{\epsilon} \in \left[\sigma_{met} , \sigma_{A} , \sigma_{B} , \sigma_{C},\sigma_{U}  \right].
\end{equation}

By allowing noisy observations to be controlled by the nugget variability, all observations can be written in terms of a common mean function and a common spatial process, thereby keeping the parameter space small for efficient inference. In Appendix \ref{PertubationEffect}, the effect of altering the noise parameter on predictions is illustrated with an example. 

\subsection{Parameter Inference: SPDE approach}
\label{sectionInference}

\begin{figure}[!h]
\centering
\includegraphics[width=0.7\linewidth]{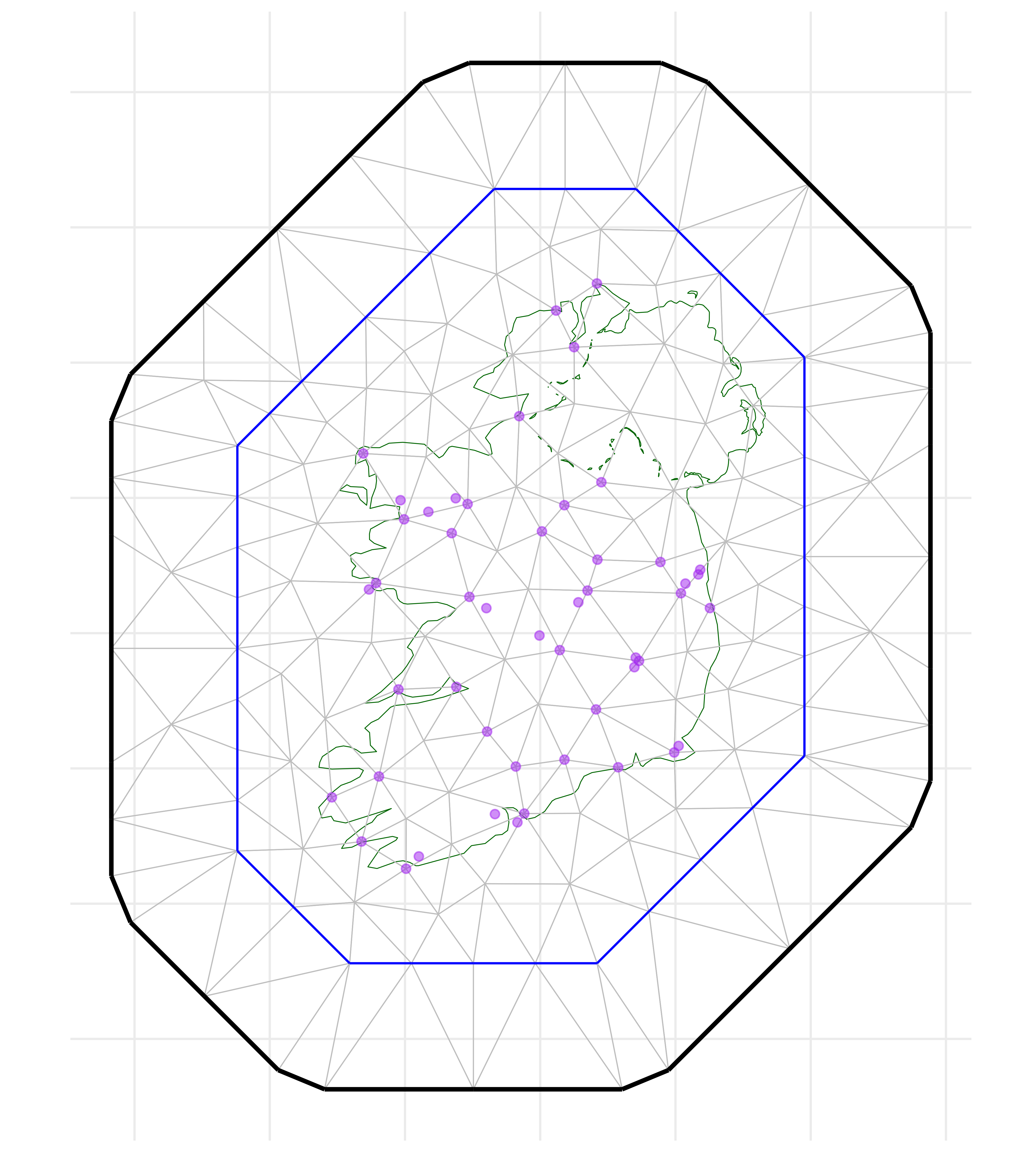}
\caption{\label{fig:IrelandMesh}The mesh used to solve the SPDE over Ireland. The mesh is divided into two zones: a higher resolution inner area containing the spatial locations of interest, and a coarser outer zone to reduce boundary effects. This mesh contains 171 nodes.}
\end{figure}

Model parameters are inferred using the Integrated Nested Laplace Approximations (INLA) framework~\citep{rue2009approximate}. \citet{lindgren2011explicit} showed that Gaussian Markov random fields (GMRFs) can be used to approximate a spatial Gaussian random field by solving a stochastic partial differential equation (SPDE) over a discretised mesh. This allows the latent spatial field to be represented with a sparse precision matrix \( Q \), significantly reducing computational cost. The spatial SPDE is given by:

\begin{equation}
    \tau\left(\kappa^{2} - \Delta \right)^{\alpha/2}\xi\left(s,t \right) = \mathcal{W}\left(s,t\right),
\end{equation}
where $\tau$ is the local precision parameter, $\kappa$ and $\alpha$ are scale and shape parameters. $\xi$ is a spatially correlated random field with Matérn covariance, and $\mathcal{W}$ is white noise. 
To implement the INLA SPDE model, the spatial domain is represented by a triangulated mesh over the study region. This mesh defines local neighbourhoods, resulting in a sparse structure for the precision matrix \( Q \). Figure~\ref{fig:IrelandMesh} shows the mesh used in this study over Ireland. It consists of a finer inner mesh covering the observation locations and a coarser outer mesh to mitigate boundary effects. The mesh contains 171 nodes. Increasing mesh resolution provided little gain in predictive performance but significantly increased computational cost.

To capture spatio-temporal dynamics, temporal dependence is added via an autoregressive model~\citep{cameletti2013spatio}. Specifically, the latent process \( z(s, t) \) evolves according to a first-order autoregressive (AR(1)) structure:

\begin{equation}
\label{AR1Spatial}
    z\left(s,t \right) = \rho z\left(s,t-1 \right) + \xi\left(s,t\right) , 
\end{equation}

where $\rho$ is the temporal correlation. $\xi\left(s,t\right)$ is a spatially correlated innovation that is independent in time and described by the spatial SPDE. This demonstrates how the hourly increments can be expressed as a spatially correlated innovation process. %In INLA, this structure is implemented using the \textit{ar1} temporal model.

One of INLA's strengths is its efficient hyperparameter estimation. INLA employs a nested approach, where the latent field and hyperparameters are inferred simultaneously through a Laplace approximation. The likelihood is approximated by integrating out the random effects (the latent field) to estimate the posterior distribution of the hyperparameters. INLA iteratively maximizes this posterior distribution using numerical optimization methods, updating both the latent field and hyperparameters \citep{rue2009approximate}.

Another advantage of INLA is that the posterior distribution of any missing values is obtained automatically as part of the model fitting process \citep{gomez2020inla_ch12}. Although the meteorological stations in our dataset contain observations at all time points, the PWS data include gaps. While handling missing data is not a primary focus of our analysis, the model nonetheless returns posterior predictions for missing PWS values, conditional on the observations from all stations.
 
To apply the INLA framework to our model with multiple observation sources, we implement it as a coregionalisation model, where different observations have a linear dependence on a shared latent field. Inference is then performed jointly over all sources. Further details on the INLA implementation of such models can be found in \citet{krainski2018chapter3}.

\subsection{Prior Specification}
For the IGP model, there are seven hyperparameters: the spatial effective range $\phi$, the standard deviation of the spatial process $\sigma_{z}$, and five parameters controlling the nugget (independent noise) for each measurement group, as specified in Equation \eqref{eq:sigmaparams}. The AR(1) model contains an additional parameter, $\rho$, which controls the temporal correlation. 

To specify informative priors for the effective range and spatial standard deviation, we use penalized complexity priors ~\citep{fuglstad2019constructing}, which allows priors to be constructed based on tail probabilities. For the effective range, the prior is defined as:
\begin{equation} P\left(\phi < 200 \right) = 0.3, \end{equation}
indicating a 30\% probability that the effective range is less than 200 km. Figure \ref{fig:SpearmanDistance} suggests that the spatial correlation decays slowly, with the potential for correlation over long distances.

For the standard deviation of the spatial process, the prior is specified as:
\begin{equation} P\left(\sigma_z > 0.75 \right) = 0.5, \end{equation}
reflecting the marginal standard deviation across all wind speeds.

Although INLA internally parametrises the independent noise in terms of precision ($\tau = 1/\sigma_{\epsilon}^2$), we specify the prior directly in terms of $\sigma_{\epsilon}$. When no prior information is available for the independent noise in each group, a common uninformed prior is used for each of the five noise parameters in Equation \eqref{eq:sigmaparams}:
\begin{equation} P\left(\sigma_\epsilon > 0.75 \right) = 0.1, \end{equation}
implying a 10\% probability that each $\sigma_\epsilon$ exceeds a value of 0.75.

If prior knowledge of the noise at meteorological stations is known, a more informed prior could be used for $\sigma_{met}$, however in our models common uninformed priors are used for each $\sigma_\epsilon$. 

For the $\rho$ parameter in the AR(1) model, analysis shows wind speeds show a temporal trend, therefore our prior will assume strong temporal correlation:
\begin{equation} P\left(\rho > 0.8 \right) = 0.7, \end{equation}
i.e. a 70\% probability that $\rho$ is greater than 0.8.

\FloatBarrier
\subsection{Evaluation methods}
\label{sec:Evaluation}
We consider two methods for evaluating the quality of our models; a method to assess the accuracy of the posterior mean, and a method to assess the quality of probabilistic forecasts. To assess the accuracy of a point prediction, a commonly used metric is the root mean square error (RMSE) \citep{lenzi2020spatiotemporal}. Given true wind speeds $W_{s,t}$ at $N$ datapoints, and predicted wind speeds $\widehat{W}_{s,t}$, the RMSE is defined as:
\begin{equation}
    \text{RMSE} = \sqrt{\frac{\sum_{i=1}^N \left(\widehat{W}_{s_i,t_i}-W_{s_i,t_i} \right)^{2}}{N}}.
\end{equation}
This measure is useful for assessing the general accuracy of the predicted values but can be sensitive to large deviations or outliers in the data.
For comparing the quality of probabilistic forecasts, where we do not have access to a true underlying distribution, we use a measure that rewards well-calibrated predictions, i.e., predicted quantiles should correspond to observed frequencies ~\citep{gneiting2007probabilistic}. A commonly used metric for this purpose is the continuous ranked probability score (CRPS). CRPS is a proper scoring rule, meaning it incentivizes truthful probabilistic forecasting by penalizing discrepancies between the predicted cumulative distribution function (CDF) and the empirical observations \citep{gneiting2007probabilistic}. This encourages both accurate calibration, where predicted probabilities match observed frequencies, and sharpness, where the predictions are as precise as possible. Given a probabilistic forecast with CDF denoted as F, and empirical observations $y_{obs}$, the average CRPS for a forecast is defined as follows:
\begin{equation}
\text{CRPS} = \frac{1}{N} \sum_{i=1}^{N} \int_{-\infty}^{\infty} \left( F_i(y) - \mathbb{I}(y \geq y_{\text{obs}, i}) \right)^2 dy,
\end{equation}
for $N$ datapoints, where $\mathbb{I}(y \geq y_{\text{obs}, i})$ is an indicator function equal to 1 if $y \geq y_{\text{obs}}$, and 0 otherwise. For practical implementation the integral can be discretized on $M$ intervals of equal size:

\begin{equation}
\text{CRPS}_{\text{avg}} = \frac{1}{N} \sum_{i=1}^{N} \sum_{j=1}^{M} \left( F_i(y_j) - \mathbb{I}(y_j \geq y_{\text{obs}, i}) \right)^2 \Delta y
\end{equation}

\subsection{Space-time Simulation Studies}
\label{Simulation}
To evaluate the effectiveness of incorporating separate noise parameters in our INLA implementation, we simulate a space-time process using the actual locations of meteorological stations and PWS from our Irish wind speed dataset. Data are simulated across 100 time points, maintaining the same grouping of stations into meteorological and PWS. We divide the PWS into two groups: PWS-1, containing stations listed as A, B, and C in Table~\ref{tab:station_class}, and PWS-2, containing stations listed under U (a total of seven).

We simulate data from a separable spatio-temporal Gaussian process with hyperparameters chosen to resemble the empirical properties of Irish wind speeds. A constant mean is assumed across all space-time points. An independent Gaussian noise is added to each observation, with a lower standard deviation assigned to meteorological stations to reflect higher-quality observations, and a higher standard deviation for PWS-1 to represent noisier data. For PWS-2, we simulate random normal data uncorrelated with the other stations to test the model's ability to detect and mitigate the contribution of poor-quality data.

To assess parameter inference and predictive performance, we compare three modelling strategies: \begin{itemize} \item \textbf{1}: Using only meteorological stations for fitting and prediction. \item \textbf{2}: Including meteorological and PWS data, assuming a single noise parameter for all stations. \item \textbf{3}: Including all station types, with separate noise parameters for meteorological, PWS-1, and PWS-2 data. \end{itemize}

Each model is fit using both the IGP and AR(1) covariance structures for the latent field. To assess predictive accuracy, we perform a leave-one-out cross-validation at each of the 23 meteorological station locations. At each test site, the remaining stations are used to predict wind speeds at all 100 time points.

To investigate the impact of noise in PWS stations, we repeat the simulation across a range of values for the PWS-1 independent noise standard deviation, \(\sigma_{\text{PWS}}\). The fixed parameters used in the simulation are summarised in Table~\ref{tab:simparams}.
\begin{table}[ht]
\centering
\caption{Simulation parameters used to generate spatio-temporal wind speed data.}
\label{tab:simparams}
\begin{tabular}{ |c|l|c| }  
 \hline
 \textbf{Parameter} & \textbf{Description} & \textbf{Value}  \\
 \hline
 $\phi$ & Effective spatial range (km) & 200 \\ 
 $\sigma_{\text{z}}$ & Standard deviation of the latent spatial field & 0.7  \\ 
 $\sigma_{\text{Met}}$ & Std. dev. of independent noise for \textit{meteorological} stations & 0.2 \\ 
 $\sigma_{\text{PWS}}$ & Std. dev. of independent noise for PWS stations & [0.3, 0.5] \\ 
 $\rho$ & AR(1) temporal correlation parameter & 0.8 \\
 \hline
\end{tabular}
\end{table} 

The accuracy results for separate simulation studies are included in Table \ref{tab:Simperformance2groups}. The estimated values of the noise hyperparameters with \(\sigma_{\text{PWS}-1}\) are included in Table \ref{tab:parameters_ar1}. The estimates of other hyperparameters are included in Table \ref{tab:parameters} in Appendix~\ref{app:Simulation_Tables}. 

\begin{table}[h!]
\centering
\begin{tabular}{|c|cc!{\vrule width 1.2pt}cc!{\vrule width 1.2pt}cc|}
\hline
\multicolumn{7}{|c|}{\textbf{Independent Replicates}} \\  
\hline
\multirow{2}{*}{Noise Level ($\sigma_{\text{PWS-1}}$)} & \multicolumn{2}{c|}{Reliable Data Only} & \multicolumn{2}{c|}{All Data (Constant Params)} & \multicolumn{2}{c|}{All Data (Different Variance)} \\
\cline{2-7}
 & RMSE & CRPS & RMSE & CRPS & RMSE & CRPS \\
\hline
0.3 & 0.46 & 0.26 & 0.44 & 0.25 & 0.40 & 0.22 \\
0.4 & 0.46 & 0.26 & 0.45 & 0.25 & 0.41 & 0.23 \\
0.5 & 0.46 & 0.26 & 0.46 & 0.26 & 0.43 & 0.24 \\
\hline
\multicolumn{7}{|c|}{\textbf{AR(1) Process}} \\  
\hline
\multirow{2}{*}{Noise Level ($\sigma_{\text{PWS-1}}$)} & \multicolumn{2}{c|}{Reliable Data Only} & \multicolumn{2}{c|}{All Data (Constant Params)} & \multicolumn{2}{c|}{All Data (Different Variance)} \\
\cline{2-7}
 & RMSE & CRPS & RMSE & CRPS & RMSE & CRPS \\
\hline
0.3 & 0.47 & 0.26 & 0.43 & 0.24 & 0.39 & 0.21 \\
0.4 & 0.47 & 0.26 & 0.43 & 0.24 & 0.40 & 0.22 \\
0.5 & 0.47 & 0.26 & 0.44 & 0.25 & 0.41 & 0.23 \\
\hline
\end{tabular}
\caption{Performance Metrics for Different Models and Noise Levels}
\label{tab:Simperformance2groups}
\end{table}

\begin{table}[h!]
\centering
\begin{tabular}{|c|c|c|c|}
\hline
Noise Level ($\sigma$) & Reliable Data Only ($\hat{\sigma}_{\epsilon}$) & \makecell{All Data (Constant Params) \\($\hat{\sigma}_{\epsilon}$)} & \makecell{All Data (Different Variance) \\($\hat{\sigma}_{\text{Met}}$, $\hat{\sigma}_{\text{PWS-1}}$,$\hat{\sigma}_{\text{PWS-2}}$)} \\
\hline
\multicolumn{4}{|c|}{\textbf{IGP Process}} \\
\hline
0.5 & 0.26 & 0.46 & (0.28, 0.52 , 0.79) \\
\hline
\multicolumn{4}{|c|}{\textbf{AR(1) Process}} \\
\hline
0.5 & 0.23 & 0.43 & (0.25,0.51, 0.70) \\
\hline
\end{tabular}
\caption{Estimated Parameter Values: Independent Noise Standard Deviation}
\label{tab:parameters_ar1}
\end{table}

These results show that when noisy data is included and assumed to come from a single homogeneous observation process, predictive error improves slightly or remains the same, as informative stations are offset by low-quality stations. However, using multiple nugget variances correctly identifies the differing noise levels associated with each group: the independent noise parameter for the poor-quality data group inflates, effectively negating its contribution to predictions. The inclusion of PWS-1 data leads to improved accuracy in both RMSE (up to 13\% reduction) and CRPS (up to 15\% reduction), highlighting increased accuracy, but also better assessment of prediction variance. The model also correctly estimates the $\sigma_{\text{PWS-1}}$ parameter associated with these observations. 

 Another result of note is that, although the data were simulated from a separable space-time model, and the AR(1) model accurately captures the covariance parameters, there is little difference in predictive accuracy between the AR(1) and IGP models. We discuss this finding further in Appendix~\ref{app:ar1vsIGP}. We also noticed that the IGP model was substantially faster to fit than the AR(1). Therefore, in applications where the goal is to interpolate a historical time series over a long period, an IGP model may be sufficient. Nevertheless, we still examine the accuracy of both approaches on our real wind dataset.

\FloatBarrier
\section{Results}
\label{SectionResults}

\subsection{Spatial Prediction Results}
We apply our models to the combined dataset of meteorological stations and bias-corrected PWS. To evaluate the impact of covariates and the inclusion of PWS, we compare several models of increasing complexity under both IGP and AR(1) covariance structures. We adopt a leave-one-station-out cross-validation (LOSOCV) strategy, in which one meteorological station is sequentially excluded from model training and used solely for evaluation. Each excluded station is predicted at all observed time points using the remaining stations, focusing on spatial interpolation rather than temporal forecasting. For each test station, we predict hourly wind speeds over a six-month period from June 2024 to November 2024 (4,392 time points). This validation is repeated for all 23 meteorological stations.

Each model implemented is described in Table \ref{tab:model_comparison}. The most complex models (5-7) are of the form described in Equation \eqref{eq:finalmodel}. The accuracy of each model is compared using the RMSE and CRPS metrics defined in Section \ref{sec:Evaluation}. The posterior mean at each location for each timepoint is used for the RMSE calculation. For the CRPS, we assume a Gaussian predictive distribution, using the posterior mean and standard deviation. The RMSE is calculated with respect to the true wind speeds, while the CRPS is calculated on the square root scale, consistent with the Gaussian assumption and simplifying the calculation. The results are shown in Table~\ref{tab:model_comparison}.

\begin{table}[h]
\centering
\begin{tabular}{ |c|c|c|c| }  
 \hline
   \textbf{Model \#} & \textbf{Model Description} & \textbf{RMSE} & \textbf{CRPS}\\
 \hline
 \multicolumn{4}{|c|}{\textbf{IGP Process}} \\
 \hline
 1 & Meteorological Stations only, constant mean &  1.78 & 0.22  \\ 
 2 & Meteorological Stations only, with covariates & 1.35 & 0.18 \\
 3 & Met and PWS (No bias correction) - constant mean & 2.51 & 0.39\\
 4 & Met and PWS (bias correction), constant mean & 1.63 & 0.2 \\
 5 & Met and PWS (bias correction), with covariates, single nugget variance  & 1.28 & 0.16 \\
 6 & Met and PWS (bias correction), with covariates, group specific variance & 1.28 & 0.16 \\
 \hline
 \multicolumn{4}{|c|}{\textbf{AR(1) Process}} \\
 \hline
 7 & Met and PWS (bias correction), with covariates, group specific variance & 1.90 & 0.30 \\
\hline
\end{tabular}
\caption{Comparison of model performance using RMSE and CRPS.}
\label{tab:model_comparison}
\end{table}

Results in Table~\ref{tab:model_comparison} show in the first place that the AR(1) model provides worse accuracy compared to the case when independent replicates in time are used, with a RMSE of 1.9 versus 1.28. This indicates that the AR(1) structure may not be suitable for the Irish wind speed data. We examine in Appendix \ref{app:ar1vsIGP} the potential reasons for this poor performance, and will focus in the following on the results from the independent replicates approach.

Amongst IGP models (1-6), the model with the poorest performance (3) used a constant mean with uncorrected PWS, yielding significantly higher errors compared to the same model using only Meteorological stations (1), with a RMSE of 2.51 versus 1.78. This highlights that raw wind speed measurements from PWS contribute little to model accuracy without transformation or correction.

The comparison of accuracy results between the model 4 and model 5 shows that including covariates, especially the mean derived from reanalysis data and distance to the sea, significantly improved model accuracy with RMSE decreasing from 1.63 to 1.28. Although other commonly used covariates, such as altitude or orography, were not explicitly included, we assume these are accounted for within the reanalysis output, which utilizes physical models of topography to simulate wind speeds.

Finally, the most accurate models (5-6) are the ones incorporating both bias-corrected stations and reanalysis-derived covariates for the mean function. Including the bias-corrected weather stations reduced the RMSE from 1.35 to 1.28, a 5.2\% reduction, and decreased the CRPS from 0.18 to 0.18, representing a 10.5\% improvement compared to using only meteorological stations. It is however surprising that both models perform similarly despite the more complex and allegedly more realistic structure of measurement errors in model 6. To investigate the reason for this similar level of performance, Table~\ref{tab:single_variance} and Table~\ref{tab:group_specific_variance} display the hyperparameters of the two models fit in INLA. The credible intervals for the hyperparameters are notably narrow, likely due to the large volume of data. When the model was fitted to shorter time periods as an experiment, the intervals became noticeably wider, suggesting that data volume influences parameter uncertainty. See Appendix \ref{app:IntervalvsTime} for the effect the number of timepoints has on the confidence intervals of the spatial parameters.
  
\begin{table}[h]
\centering
\begin{tabular}{ |c|c|c|c| }  
 \hline
   \textbf{Parameter} & \textbf{Value (Quantile Bias Correction)} & \textbf{2.5 Quantile} & \textbf{97.5 Quantile} \\
 \hline
 $\phi$ (km) & 526  & 513 &  535\\ 
 $\sigma_{\text{z}}$ & 0.326 & 0.321 & 0.330 \\ 
 $\sigma_{\epsilon}$ & 0.282 & 0.281 & 0.284 \\ 
 \hline
\end{tabular}
\caption{Estimated parameters for model 5, which has a single variance parameter for observations.}
\label{tab:single_variance}
\end{table}

\begin{table}[h]
\centering
\begin{tabular}{ |c|c|c|c| }  
 \hline
   \textbf{Parameter} & \textbf{Value (Quantile Bias Correction)} & \textbf{2.5 Quantile} & \textbf{97.5 Quantile} \\
 \hline
 $\phi$ (km) & 491  & 482 &  500\\ 
 $\sigma_{\text{z}}$ & 0.336 & 0.331 & 0.340 \\ 
 $\sigma_{\text{Met}}$ & 0.259 & 0.258 & 0.261 \\ 
 $\sigma_{\text{A}}$ & 0.285 & 0.277 & 0.294 \\ 
 $\sigma_{\text{B}}$ & 0.297 & 0.294 & 0.300 \\ 
 $\sigma_{\text{C}}$ & 0.315 & 0.312 & 0.317 \\ 
 $\sigma_{\text{U}}$ & 0.274 & 0.271 & 0.277 \\ 
 \hline
\end{tabular}
\caption{Estimated parameters for Model 6, which includes separate variance parameters for each observation group.}
\label{tab:group_specific_variance}
\end{table}

Results in Table~\ref{tab:group_specific_variance} show that in opposition with our initial hypothesis suggesting that lower wind grades should have higher observation noise, all groups show similar noise parameters. In addition, comparing Tables~\ref{tab:single_variance} and~\ref{tab:group_specific_variance} shows the level of noise in each of the station groups is similar to the overall noise when a single parameter is used in model 5, which explains the similar performance of the two models. This result may be due to preprocessing steps such as bias correction and filtering out low-correlation sites, which results in corrected PWS data being of similar accuracy to meteorological stations, as well as ensuring PWS stations have similar levels of variability to meteorological stations.

To test the model’s ability to detect unreliable groups of wind measurements, we fit a version where the `unknown rating' group (U) remains uncorrected (8 of 49 stations) and reiterated the comparison between models 5 and model 6 in terms of model accuracy (Table~\ref{tab:nugget_comparison}) and parameter estimates (Tables~\ref{tab:single_variance_2} and~\ref{tab:group_specific_variance_2}).

\begin{table}[h]
\centering
\begin{tabular}{ |c|c|c|c| }  
 \hline
  & \textbf{Model} & \textbf{RMSE} & \textbf{CRPS} \\
 \hline
 5 & Met and PWS (bias correction), with covariates, single nugget variance  & 1.53 & 0.20 \\
 6 & Met and PWS (bias correction), with covariates, group-specific variance & 1.32 & 0.18 \\
 \hline
\end{tabular}
\caption{Comparison of model 5 and 6 where the unknown (U) station group has been deliberately left uncorrected.}
\label{tab:nugget_comparison}
\end{table}

\begin{table}[h]
\centering
\begin{tabular}{ |c|c|c|c| }  
 \hline
   \textbf{Parameter} & \textbf{Value (Quantile Bias Correction)} & \textbf{2.5 Quantile} & \textbf{97.5 Quantile} \\
 \hline
 Effective Range (km) & 581  & 570 &  596\\ 
 $\sigma_{\text{z}}$ & 0.31 & 0.30 & 0.31 \\ 
 $\sigma_{\epsilon}$ & 0.44 & 0.44 & 0.45 \\ 
 \hline
\end{tabular}
\caption{Estimated parameters for model 5 with a single variance parameter for observations where the Unknown (U) group is deliberately left uncorrected. In this $\sigma_{\epsilon}$ has grown and is larger than $\sigma_{\text{z}}$.}
\label{tab:single_variance_2}
\end{table}

\begin{table}[h]
\centering
\begin{tabular}{ |c|c|c|c| }  
 \hline
   \textbf{Parameter} & \textbf{Value (Quantile Bias Correction)} & \textbf{2.5 Quantile} & \textbf{97.5 Quantile} \\
 \hline
 Effective Range (km) & 579  & 568 &  595\\ 
 $\sigma_{\text{z}}$ & 0.339 & 0.335 & 0.343 \\ 
 $\sigma_{\text{Met}}$ & 0.27 & 0.264 & 0.269 \\ 
 $\sigma_{\text{A}}$ & 0.27 & 0.258 & 0.263 \\ 
 $\sigma_{\text{B}}$ & 0.25 & 0.237 & 0.246 \\ 
 $\sigma_{\text{C}}$ & 0.27 & 0.264 & 0.272 \\ 
 $\sigma_{\text{U}}$ & 0.89 & 0.87 & 0.90 \\ 
 \hline
\end{tabular}
\caption{Estimated parameters for model 6 with separate variance parameters for each observation group where the Unknown (U) group is left uncorrected. In this model $\sigma_{\text{U}}$ as grown large, reducing the contribution of the U group to predictions.}
\label{tab:group_specific_variance_2}
\end{table}

The results in Table~\ref{tab:nugget_comparison} show that model 5, characterized by a single likelihood for measurement errors, is strongly impacted by noisy data. The measurement error parameter ($\sigma_{\epsilon}$) increases from 0.26 to 0.44 and accuracy metrics degrade significantly (RMSE increases from 1.28 to 1.53 and CRPS from 0.16 to 0.20). In contrast, the multiple likelihood approach of model 6 is subject to a limited performance decrease (RMSE increases from 1.28 to 1.32 and CRPS goes from 0.16 to 0.17). In addition, the noise parameter $\sigma_{U}$ inflates significantly in this setting (Table~\ref{tab:group_specific_variance_2}), effectively mitigating the influence of data from poor-quality stations. 

The above suggests that while preprocessing successfully removed low-quality stations, the multiple likelihood approach of model 6 may still be valuable in automated models where data quality could degrade or station setups might change, leading to different distributions of measurement errors. Additionally, this approach is useful in scenarios where data is only available at a single time point or a limited number of time points, preventing correlation-based quality checks or quantile transformations. In such cases, the use of multiple variance parameters could be employed to evaluate the reliability of different groups.

As a last comparison, Table~\ref{tab:comp_model6_ERA} evaluates our best model (model 6) against estimated hourly wind speeds from ERA5, a product widely used in the wind industry for assessing wind resource potential. %Previous studies have identified ERA5 as the most accurate reanalysis product for estimating hourly wind speeds in Ireland \citep[e.g.][]{doddy2021reanalysis}. 
ERA5 provides weather variables on a 0.25° × 0.25° grid and wind speeds were interpolated to meteorological station locations using ordinary Kriging \citep{diggle1998model}. We again use Ireland's 23 meteorological stations to test accuracy. The time period is a six-month period from June 2024 to November 2024, and the error is based on the hourly wind speed predictions.

\begin{table}[h]
\centering
\begin{tabular}{ |c|c|c|c| }  
 \hline
   & Model & RMSE \\
 \hline
 6 & Met and PWS (bias correction), with covariates, group specific variance & 1.28 \\
 & ERA5 & 1.37 \\
 \hline
\end{tabular}
\caption{Comparison of wind speed prediction accuracy for model 6 and ERA reanalysis.}
\label{tab:comp_model6_ERA}
\end{table}

Results in Table~\ref{tab:comp_model6_ERA} show that our statistical model offers an improvement over ERA5 in the 6-month test period, with a RMSE of 1.28 versus 1.37. While the improvement is small (6.5\%), and there are stations where ERA5 is more accurate, this approach offers the advantage of being available in real-time. This improves the potential for short-term forecasts, an important tool for grid management. In contrast, ERA5 is only available after a five-day delay. The statistical approach also provides robust uncertainty quantification, which is not readily available with reanalysis products.

\begin{figure}[H] 
    \centering
    % First row
    \begin{subfigure}[b]{0.45\textwidth}
        \includegraphics[width=\textwidth]{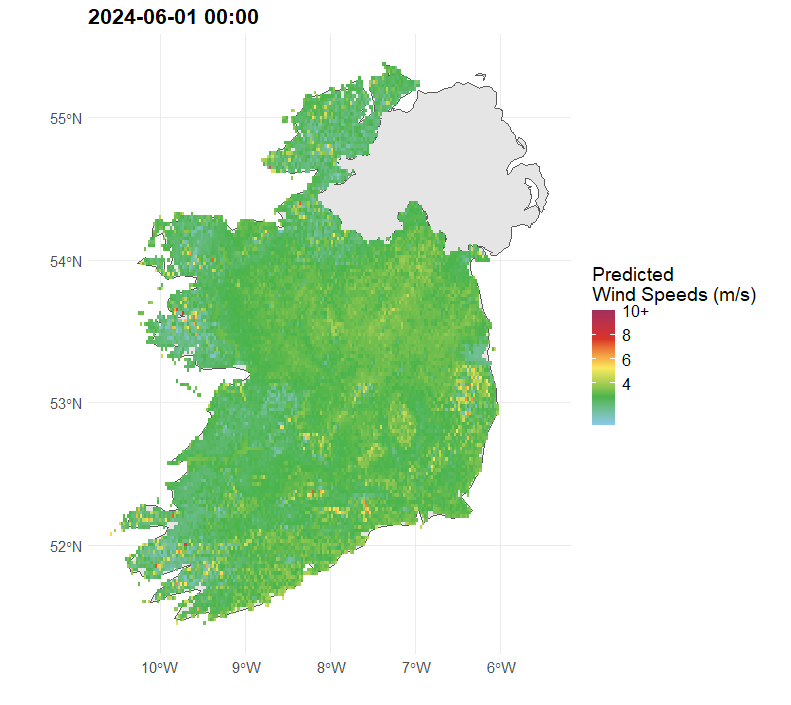}

        \label{fig:plot1_prediction}
    \end{subfigure}
    \hfill
    \begin{subfigure}[b]{0.45\textwidth}
        \includegraphics[width=\textwidth]{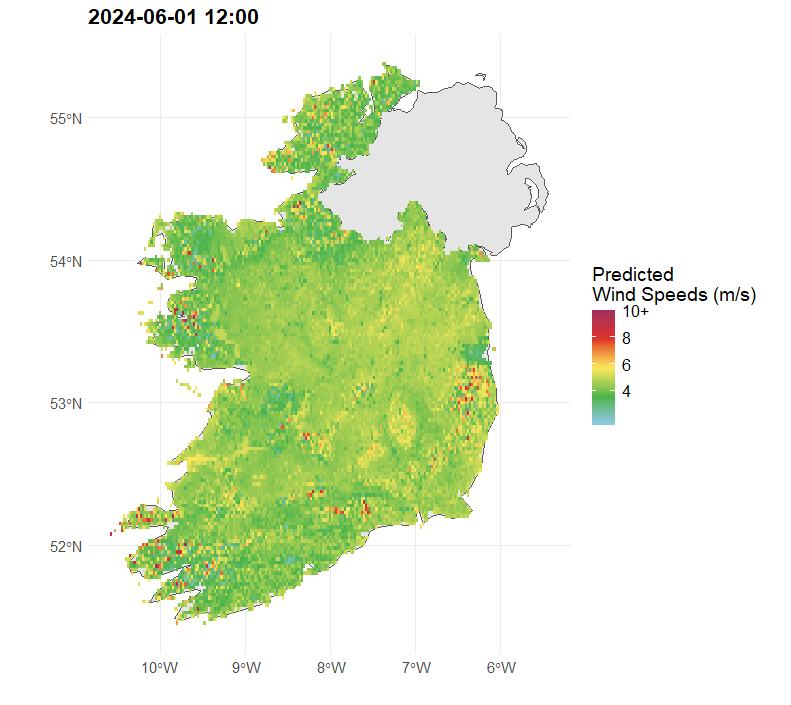}

        \label{fig:plot2_prediction}
    \end{subfigure}
    
    % Second row
    \vskip\baselineskip
    \begin{subfigure}[b]{0.45\textwidth}
        \includegraphics[width=\textwidth]{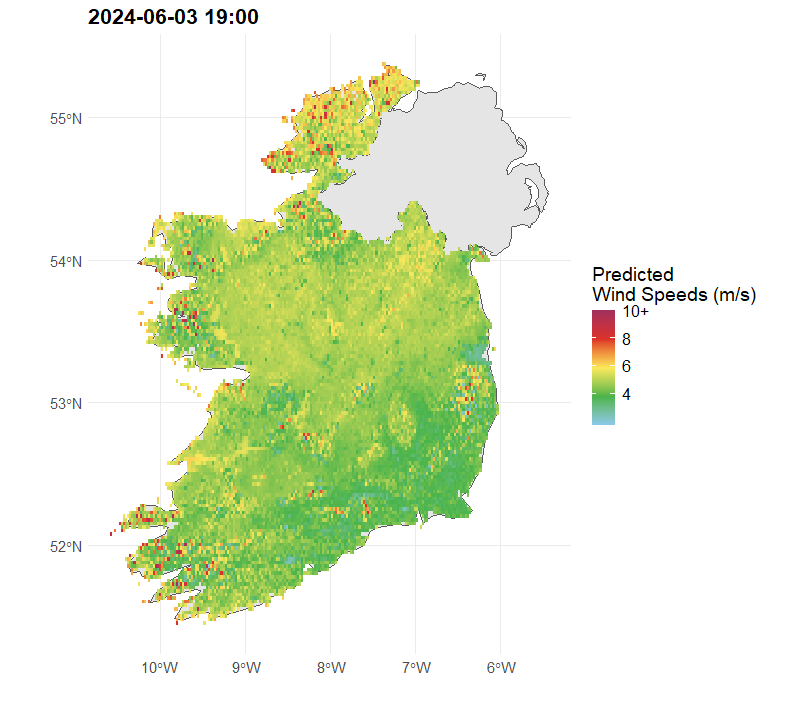}
 
        \label{fig:plot3_prediction}
    \end{subfigure}
    \hfill
    \begin{subfigure}[b]{0.45\textwidth}
        \includegraphics[width=\textwidth]{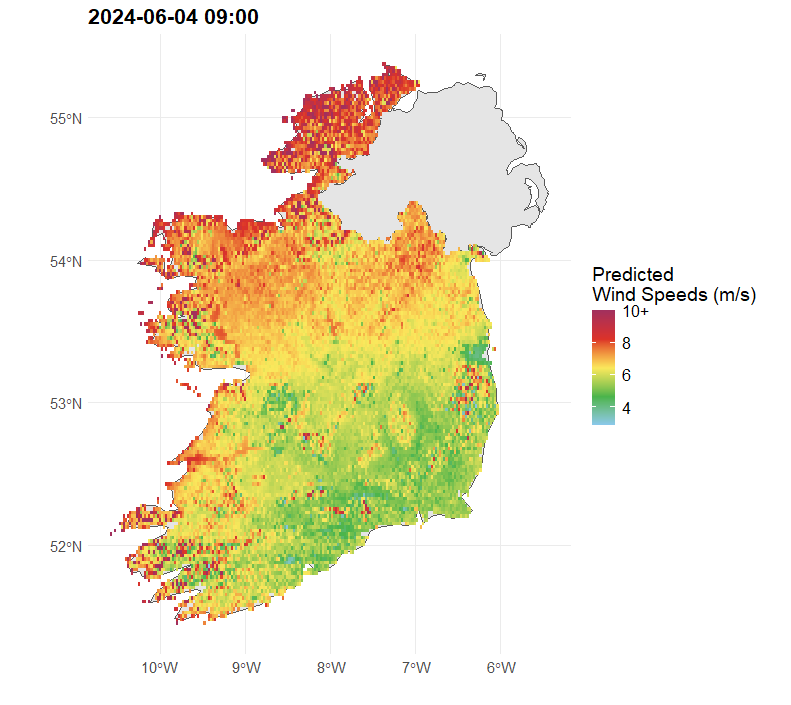}

        \label{fig:plot4_prediction}
    \end{subfigure}
    
    \caption{Predicted wind speeds using model 6 at a sample of four time points on a $0.025^\circ\times0.025^\circ$ grid. We choose four time points that have different wind profiles (both low and high wind speeds). Fine scale local variations in wind speeds are driven by the covariates derived from reanalysis. }
    \label{fig:PredictionGrid}
\end{figure}

The wind fields estimated by the best model (model 6) are illustrated in Figure~\ref{fig:PredictionGrid} for four time points in June 2024, which shows typical wind patterns across Ireland that emerge from the best model.

\subsection{Performance under Extreme Wind Conditions}
\label{sec:extremes}
We also assess how accurately our model predicts extreme wind speeds. In the context of renewable energy, extreme wind speeds lead to wind curtailment and can also lead to damage to wind turbines\citep{hering2010powering}. Outside of renewable energy applications, extreme wind speeds can potentially damage built structures and pose a risk to life. To assess the accuracy of predicting extreme wind speeds we compare the RMSE, bias (predicted - actual) and correlation of wind speeds for the top 1\%, 2.5\% and 5\% of wind speeds. Table \ref{tab:comp_extremes} contains the metrics for our final model 6, and compared to ERA5. \\

\begin{table}[h!]
\centering
\begin{tabular}{|c|l|c|c|c|}
\hline
\textbf{Percentile} & \textbf{Model} & \textbf{RMSE} & \textbf{Bias} & \textbf{Pearson Correlation} \\
\hline
\multirow{2}{*}{5\%} 
& Met + PWS (bias corrected), covariates, group--specific variance &  2.43 & -1.78 & 0.82 \\
& ERA5 & 2.35 & -1.27 & 0.73 \\
\hline
\multirow{2}{*}{2.5\%} 
& Met + PWS (bias corrected), covariates, group--specific variance &  2.79 & -2.15 & 0.82 \\
& ERA5 & 2.60 & -1.50 & 0.74 \\
\hline
\multirow{2}{*}{1\%} 
& Met + PWS (bias corrected), covariates, group--specific variance &  3.20 & -2.54 & 0.83 \\
& ERA5 & 2.89 & -1.80 & 0.75 \\
\hline

\end{tabular}
\caption{Predictive performance under extreme wind conditions. 
Metrics reported for the top 5\%, 2.5\% and 1\% of observed wind speeds.}
\label{tab:comp_extremes}
\end{table}

Across all thresholds, the model tends to underpredict the highest wind speeds, as seen by the negative bias values. Analysis of the underlying data shows that extreme wind speeds in Ireland are often highly localised, especially at exposed coastal stations, and are not typically observed at nearby inland stations. Because our model relies on spatial information borrowed from neighbouring sites, these isolated extremes are more difficult to capture. In contrast ERA5, although it exhibits lower overall accuracy, performs slightly better in the upper tail. ERA5 gains from its physics based model, which capture storm conditions, as well as modelling wind fields over the surrounding ocean. As more observations are included (e.g., top 5\% rather than top 1\%), the performance gap narrows, and our model exhibits consistently higher correlation with the observed wind speeds.

\subsection{Performance of AR(1) versus Independent Gaussian Processes}
\label{app:ar1vsIGP}
Across both the simulation study and the wind speed dataset is that the independent Gaussian process (IGP) model performs comparably to, or even outperforms the AR(1) model, despite the latter’s theoretical advantage in capturing temporal correlation. In the simulation study, this result can be explained by the separable spatio-temporal process used for data generation. For a separable process, the value at an unobserved location conditionally depends only on other locations at the same time point, and its own past or future values at the same location. Since the test case involves predictions at a completely unobserved location, the temporal component of the AR(1) model provides no additional information, reducing its predictive advantage over the purely spatial IGP model, see Table~\ref{tab:Simperformance2groups}.

For the wind speed data, the estimated AR(1) correlation parameter is very close to 1 ($\rho \approx 0.98$). Notably, the spatial range parameter in the AR(1) model is smaller than that of the IGP model, indicating that the AR(1) model attributes most of the observed dependence to temporal persistence rather than spatial structure (See Table \ref{tab:group_specific_variance_IGP_AR1}). Consequently, predictions at unobserved locations often tend toward the station-specific mean, as the reduced spatial range limits borrowing of information across stations. This behaviour likely explains the IGP’s superior performance—its larger spatial range allows it to leverage cross-station correlations more effectively.

\begin{figure}[H] 
    \centering
    % First row
    \begin{subfigure}[b]{0.45\textwidth}
        \includegraphics[width=\textwidth]{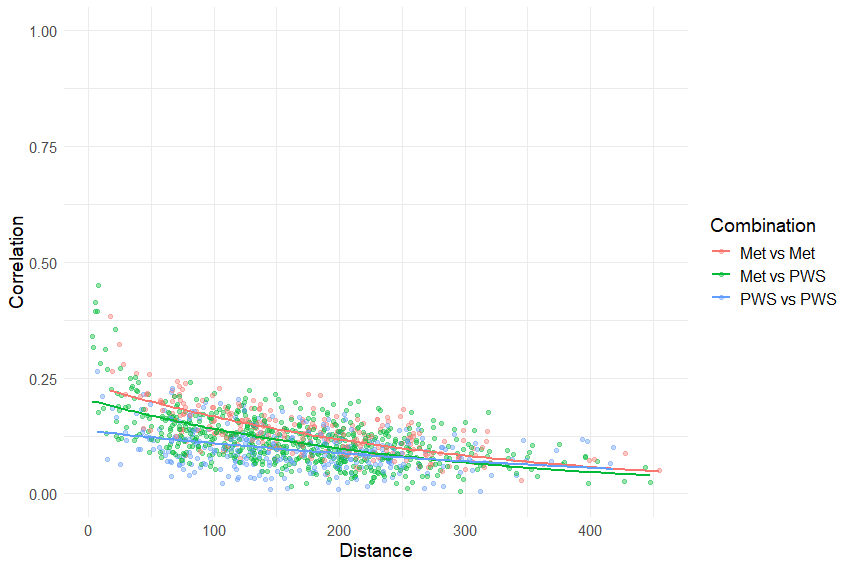}
        \caption{(a)}
        \label{fig:CorrDistanceInnovation}
    \end{subfigure}
    \hfill
    \begin{subfigure}[b]{0.45\textwidth}
        \includegraphics[width=\textwidth]{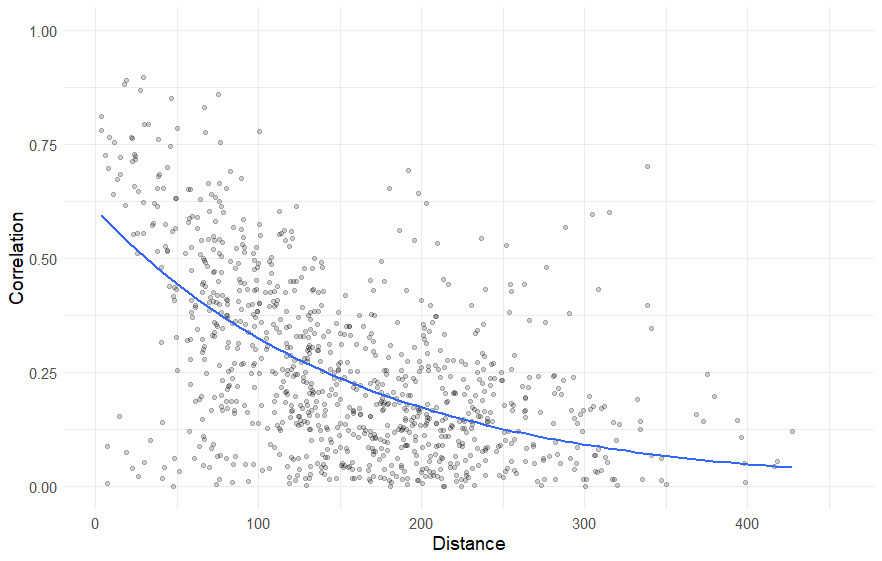}
        \caption{(b) }
        \label{fig:SimularedDistCorrDiff}
    \end{subfigure}
    
    \caption{A plot of empirical correlation against distance (in kilometres) for the hourly increments for (a) the wind speed dataset and (b) the simulated data from a true AR(1) process in Section \ref{Simulation}.}
    \label{fig:InnovationCompare}
\end{figure}

The limited range may arise from how the AR(1) parameters are inferred. The spatio-temporal process at time $t$ is expressed as \[z_{s,t}= \rho z_{s,t-1}+\xi_{s,t}, \] $\xi_{s,t}$ represents a spatially correlated innovation term between $t$ and $t+1$, which is used to infer spatial hyperparameters. Empirically, the hourly innovations in the wind data show much lower spatial correlation than expected under a true AR(1) process. In Figure \ref{fig:InnovationCompare} the correlation of hourly changes against distance is compared in both the wind dataset and the simulated data from a true AR(1) process, demonstrating how hourly changes in the wind dataset are less correlated than expected in simulated data. This mismatch likely leads the AR(1) model to underestimate the spatial range and attribute most of the structure to temporal dependence, thereby reducing its predictive performance relative to the IGP.

\begin{table}[h]
\centering
\begin{tabular}{ |c|c|c| }  
 \hline
   \textbf{Parameter} & \textbf{Parameter Value - IGP} & \textbf{Parameter Value - AR(1)}  \\
 \hline
 Effective Range (km) & 491  & 103 \\ 
 $\sigma_{\text{z}}$ & 0.336 & 0.57  \\ 
 $\sigma_{\text{Met}}$ & 0.259 & 0.123  \\ 
 $\sigma_{\text{A}}$ & 0.285 & 0.221  \\ 
 $\sigma_{\text{B}}$ & 0.297 & 0.158 \\ 
 $\sigma_{\text{C}}$ & 0.315 & 0.240 \\ 
 $\sigma_{\text{U}}$ & 0.274 & 0.162  \\ 
 $\rho$ &  & 0.97  \\ 
 \hline
\end{tabular}
\caption{Estimated parameters for Model 6 under both the IGP and AR(1) covariance structure.}
\label{tab:group_specific_variance_IGP_AR1}
\end{table}

\section{Discussion}
In this paper, we developed a spatio-temporal model that integrates both official meteorological observations and crowdsourced personal weather station (PWS) data. Our analysis shows that, although PWS tend to underestimate wind speeds compared to meteorological observations, they correlate well with nearby meteorological stations. This illustrates the trade-off between gaining increased spatial coverage and the uncertainty introduced by the lower quality of crowdsourced data. To address potential reliability issues with crowdsourced data, we implemented two key steps.

The first step involved bias correction of the crowdsourced data using reanalysis datasets, which provide gridded coverage and incorporate physical models to capture key atmospheric processes. This allowed us to estimate wind speed distributions at locations either unobserved or observed only by PWS. In this study, we used the Global Wind Atlas (GWA), but future work could explore other reanalysis datasets or model-averaging approaches to mitigate dataset-specific biases and improve correction robustness~\citep{raftery2005using}. Applying an inverse CDF transform, we used the reanalysis distributions to adjust the PWS data, yielding corrected wind speeds that more closely align with expected values.

The second step involved incorporating the crowdsourced data into a spatio-temporal model to estimate wind speeds at unobserved locations. To account for varying data reliability, we introduced a nugget variance parameter that varies by station category within the crowdsourced dataset. This controls the influence of noisy observations on the posterior mean, while preserving a manageable number of parameters. Although the empirical results showed similar variance across station categories, simulation experiments in Section \ref{Simulation} demonstrated that our approach can successfully differentiate between groups with varying levels of noise. Future work could extend this by introducing site-specific parameters or allowing variance to depend on spatially or temporally varying covariates (e.g., elevation, anemometer type, wind direction). This could more accurately capture the reliability of a crowdsourced station at a given point in time, at the expense however of a more highly parameterised model. 

Our results underscore the importance of bias correction when incorporating crowdsourced data into predictive models. Including the bias-corrected weather stations led to an 5\% reduction in error, while including non-bias-corrected stations resulted in a significant increase in error. When incorporating crowdsourced stations, our leave-one-out predictive accuracy was similar to, and in some cases slightly better than, predictions by ERA5 (6.5\% decrease in error overall). As the number of PWS increase in the future we expect the performance of models to further improve.

ERA5 is one of the most commonly used reanalysis products; however, it is only available with a five-day delay. A statistical approach, on the other hand, has the added benefit of real-time availability. One caveat to our results is that our leave-one-out stations were not part of a fully independent test set. Since we tested on meteorological stations that also inform physical weather models, which in turn are used for bias correction, the results at these sites may be more accurate than those from a truly independent test set.

Future work will explore accuracy using an independent dataset, such as wind farm data, which could offer a better test of the model’s generalisability. As wind farms typically operate on a sub-hourly level, this may require aggregating data at finer temporal resolutions. Since many PWS provide sub-hourly data, they could support the generation of high-resolution wind time series, a type of output not available from reanalysis datasets. Another area of future research is to combine reanalysis data such as ERA5 and weather station observations into a single model. While we considered them as competing models in this study, numerous studies have investigated jointly modelling station observations and gridded weather models \citep[e.g.][]{forlani2020joint,paciorek2012combining}. In forecasting applications, integrating observations with NWP has been shown to yield superior predictive accuracy compared with using NWP alone \citep{chen2021data}.

To simplify computations, we employed a stationary covariance structure. We also compared a spatial-only approach with a separable AR(1) model; however, the spatial-only approach resulted in lower errors. This suggests that an AR(1) structure does not adequately capture the space-time interaction in wind data. Although this approach does not fully represent the underlying process, previous studies have shown that stationary or separable covariance structures can significantly reduce computational costs with only a minor loss in accuracy.

Nonetheless, considerable research has focused on efficient non-stationary and non-separable models, such as nearest-neighbour Gaussian processes or SPDE approaches that model advection and diffusion in space and time to capture transport processes \citep{datta2016nonseparable,clarotto2024spde}. Future research may explore implementing these advanced models to better capture non-stationary and non-separable behaviour, which could improve short-term forecasts.

Although this application focused on wind speed, we could extend the models to incorporate both wind speed and direction \citep[e.g.][]{lenzi2020spatiotemporal,murphy2025joint}, another important value for wind energy assessment. The same bias correction and spatial modelling approach could also be applied to other weather variables where sparse official observations can be supplemented with crowdsourced data. Reanalysis products, such as ERA5 and MERRA2, which provide a wide range of variables, could similarly be used to bias-correct the empirical distribution of other weather parameters. However, extending this methodology to different variables may require adjustments, such as modifying the spatio-temporal correlation structures to better capture the unique dynamics of each variable. For example, temperature and precipitation may exhibit different correlation properties compared to wind speed, which could influence model performance. Also variables such as precipitation may require zero-inflated models.

\section*{Data Availability}

The code and data used in this study are available at: 
\href{https://github.com/EamonnO22/Spatio-Temporal-Wind-Modelling-with-Bias-Corrected-Crowdsourced-Data}{github.com/EamonnO22}

\section{Acknowledgements}
This publication has emanated from research supported in part by a grant from Taighde Éireann – Research Ireland under Grant number 18/CRT/6049. This publication has emanated from research conducted with the financial support of
the EU Commission Recovery and Resilience Facility under the Research Ireland
Energy Innovation Challenge Grant Number 22/NCF/EI/11162G. This work was partially supported by the “PHC ULYSSES” program (project number:  50252NC ), funded by the French Ministry for Europe and Foreign Affairs, the French Ministry for Higher Education and Research, Taighde Eireann – Research Ireland.
The French authors also acknowledge the financial support of the Chair Geolearning, funded by ANDRA, BNP Paribas, CCR and the SCOR Foundation for Science. 
For the purpose of Open Access, the author has applied a CC BY public copyright licence to any Author Accepted Manuscript version arising from this submission.

Data obtained from the Global Wind Atlas version 3.3, a free, web-based application developed, owned and operated by the Technical University of Denmark (DTU). The Global Wind Atlas version 3.3 is released in partnership with the World Bank Group, utilizing data provided by Vortex, using funding provided by the Energy Sector Management Assistance Program (ESMAP). For additional information: \href{https://globalwindatlas.info}{https://globalwindatlas.info}

\newpage
\appendix
\section{Empirical study of Weibull distribution for Irish wind data}
\label{app:WeibullStudy_qq}
\begin{figure}[!htbp]
\centering
\includegraphics[width=0.7\textwidth]{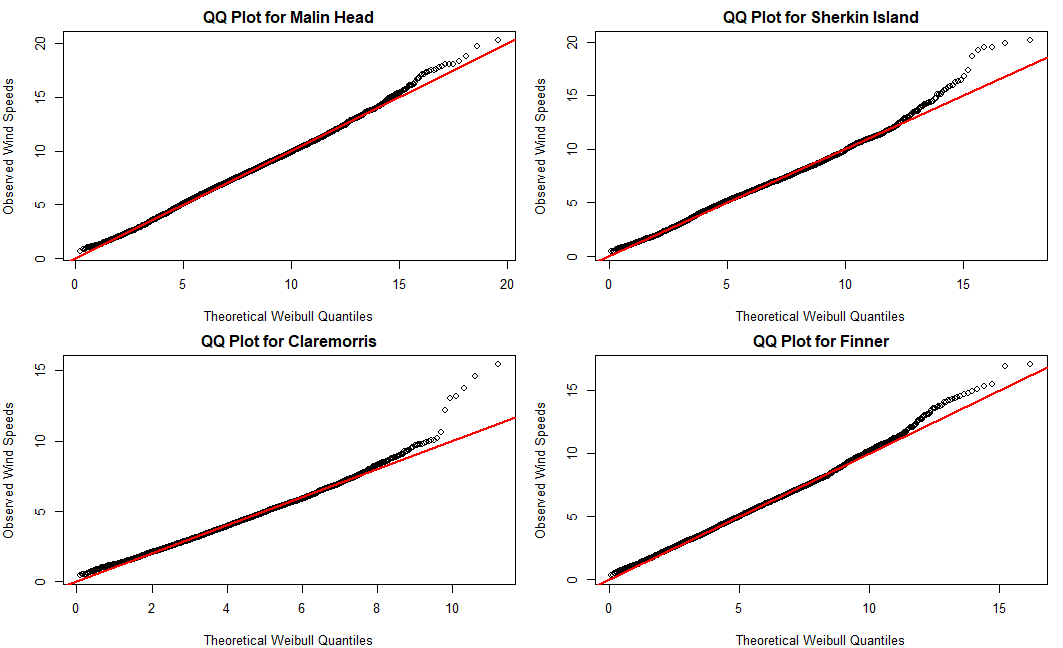}
\caption{\label{fig:QQ} Q-Q plots at four Meteorological stations, comparing theoretical to empirical Weibull quantiles. The plots demonstrate the Weibull distribution shows good agreement for most of the data, but empirical data contains more extreme values than modelled by a Weibull}
\end{figure}

\FloatBarrier
\section{Bias Correction Methods}
\label{BiasCorrection}
A potential approach is to incorporate spatially varying coefficients in the model, which allows the coefficients of the covariates to vary in space \citep{finley2020bayesian}. However, during inference these models showed a poor fit, most likely due to the relative sparsity of data (only 22 stations were used to fit these models). They had lower predictive accuracy using a leave one out error measurement when compared to spatially independent models. Therefore to incorporate the spatial patterns seen in the residuals in Figure \ref{fig:GWA_residual_maps}, distance from the sea (in kilometres) was including as a covariate. 

The second extension was to include non-linearity in the models. The reanalysis estimates, in particular for $\lambda^{GWA}_s$, shows a potential non-linear relationship with  $\lambda^{MET}_s$. To incorporate non-linearity, a Generalized Additive Model (GAM) was fit which used a smooth spline term. This was fit using the \textit{mgcv} package \citep{wood2015package}. The default smooth term, a thin plate spline was used. 

To compare our calibration models, each one was fit to 22 stations, and a prediction was performed at the $23^{rd}$ station. This was repeated 23 times to predict at all stations. Accuracy was compared using a root mean square error metric and AIC. We simplify our model by assuming that \( \lambda^{\text{MET}}_s \) and \( k^{\text{MET}}_s \) follow a normal distribution. While this assumption does not strictly hold—since these parameters are constrained to be positive—we focus only on point estimates, and in practice, all predicted values are well above zero. An alternative modelling approach for constrained parameters would be to model these parameters on the log scale. However given our covariates are on the same scale as our response, we choose not to perform any transformations.

\begin{table}[h!]
\centering
\caption{Modelling Weibull Shape Parameter}
\begin{tabular}{ |c|c|c|c| }  
 \hline
   Model & Model Structure & Estimated Model & RMSE \\
 \hline
 1 & $k = k^{GWA}$ &  & 0.30  \\ 
 2 & $k \sim  \beta_{0} + \beta_{1}k^{GWA}$ & $\hat{k} = 0.66 + 0.90k^{GWA}$ & 0.13 \\ 
 \hline
\end{tabular}
\end{table}

\begin{table}[h!]
\centering
\caption{Modelling Weibull Scale Parameter}
\begin{tabular}{ |c|c|c|c|c| }  
 \hline
   Model & Model Structure & Estimated Model & RMSE & AIC \\
 \hline
 1 & $\lambda = \lambda^{GWA}$ &  & 1.07 & \\ 
 2 & $\lambda \sim  \beta_{0} + \beta_{1}\lambda^{GWA}$ & $\hat{\lambda} = -1.5 + 1.1\lambda^{GWA}$& 0.56 & 44.6 \\
 3 & $\lambda \sim  \beta_{0} + \beta_{1}\lambda^{GWA} + \beta_{2}Dist$ & $\hat{\lambda} = -0.55 + 0.97\lambda^{GWA} -0.001 Dist$ & 0.53 & 44.2\\
 4 & $\lambda \sim  \beta_{0} + s\left(\lambda^{GWA}\right) $ & $\hat{\lambda} = 5.1 + s(\lambda^{GWA})$ & 0.55 & 44.8 \\
 5 & $\lambda \sim  \beta_{0} + s\left(\lambda^{GWA}\right) + \beta_{1}Dist$ & $\hat{\lambda} = 5.3 + s(\lambda^{GWA}) -0.01 Dist$ & 0.49 & 42.2 \\
 \hline
\end{tabular}
\end{table}

\begin{figure}[!htbp]
\centering
\includegraphics[width=0.6\textwidth]{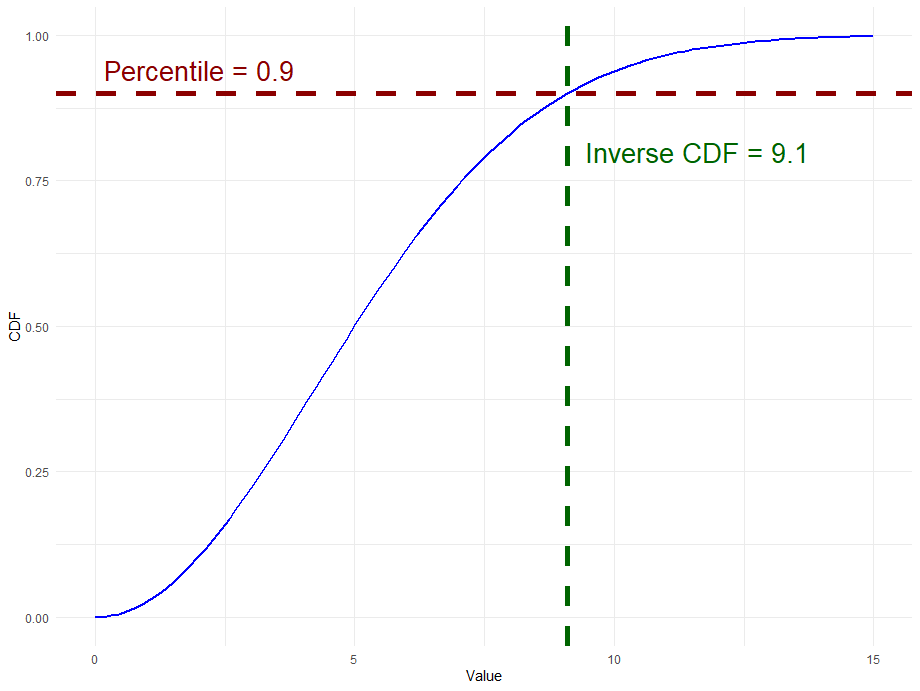}
\caption{\label{fig:InverseCDF} Illustrative example of an inverse CDF transform. An observation with a percentile of 0.9 is mapped to a corrected wind speed. The blue function denotes the CDF of a Weibull with $k = 2$ and $\lambda =6$.}
\end{figure}

To visualise the bias correction approach, Figure \ref{fig:InverseCDF} shows a sample inverse quantile transform for an arbitrary quantile and Weibull distribution. We choose a sample percentile of 0.9, representing a wind speed greater than 90\% of the raw wind speed observations, and a sample Weibull CDF with $k = 2$ and $\lambda =6$ is used (arbitrarily chosen as it reflects the realistic range of values for Irish wind speeds).
\FloatBarrier

\section{Square Root Transformation of Weibull Variable}
\label{App:sqrt_Weibull}
Let \( X \sim \text{Weibull}(k, \lambda) \), with CDF:
\[
F_X(x) = 1 - \exp\left( -\left( \frac{x}{\lambda} \right)^k \right)
\]

Define \( Y = \sqrt{X} \Rightarrow X = Y^2 \). Then:
\[
F_Y(y) = \mathbb{P}(Y \leq y) = \mathbb{P}(X \leq y^2) = F_X(y^2)
\]

Substitute:
\begin{equation*}
    F_Y(y) = 1 - \exp\left( -\left( \frac{y^2}{\lambda} \right)^k \right) 
= 1 - \exp\left( -\left( \frac{y}{\sqrt{\lambda}} \right)^{2k} \right),
\end{equation*}
which is in the form of a Weibull distribution, $Y\sim \text{Weibull}(\sqrt{\lambda},2k)$.

\section{Effect of perturbing the nugget variance on prediction.}
\label{PertubationEffect}
To highlight the effect of allowing varying nugget parameters on the predictive contribution at an unobserved location, we will visualise a toy example. Assume we have a spatial vector \textbf{z} with covariance matrix $\Sigma$, and observed data \textbf{y} with independent noise $\sigma_{y}^{2}$.
\begin{equation*}
    \textbf{Z} \sim {\cal N}\left(\textbf{0} , \Sigma + \sigma^{2}I \right)
\end{equation*}

Assuming we are interested in the posterior distribution at unobserved locations, denoted as $\textbf{Y}^{*}$, conditional on the observed data \textbf{Y}. The posterior mean and covariance of the new locations can be defined as:
\begin{equation}
    \boldsymbol{\mu}_{\mathbf{z}^{*}|\mathbf{z}} = \boldsymbol{\mu}_{\mathbf{z}^{*}} + \boldsymbol{\Sigma}_{\mathbf{z}^{*}\mathbf{z}} \boldsymbol{\Sigma}^{-1}_{\mathbf{z}\mathbf{z}} (\mathbf{z} - \boldsymbol{\mu}_{\mathbf{z}}),
\end{equation}
\begin{equation}
    \boldsymbol{\Sigma}_{\mathbf{z}^{*}|\mathbf{z}} = \boldsymbol{\Sigma}_{\mathbf{z}^{*}\mathbf{z}^{*}} - \boldsymbol{\Sigma}_{\mathbf{z}^{*}\mathbf{z}} \boldsymbol{\Sigma}^{-1}_{\mathbf{z}\mathbf{z}} \boldsymbol{\Sigma}_{\mathbf{z}\mathbf{z}^{*}}.
\end{equation}

Assuming $\mu = 0$, then our posterior mean becomes a linear combination of known values, with weights equal to $\boldsymbol{\Sigma}_{\mathbf{z}^{*}\mathbf{z}} \boldsymbol{\Sigma}^{-1}_{\mathbf{z}\mathbf{z}}$ \citep{diggle1998model}. In the stationary case, where there are common parameters across all data points, the weights are related to the distance between the prediction location and the observed value.

\begin{figure}[!h]
\centering
\includegraphics[width=0.9\linewidth]{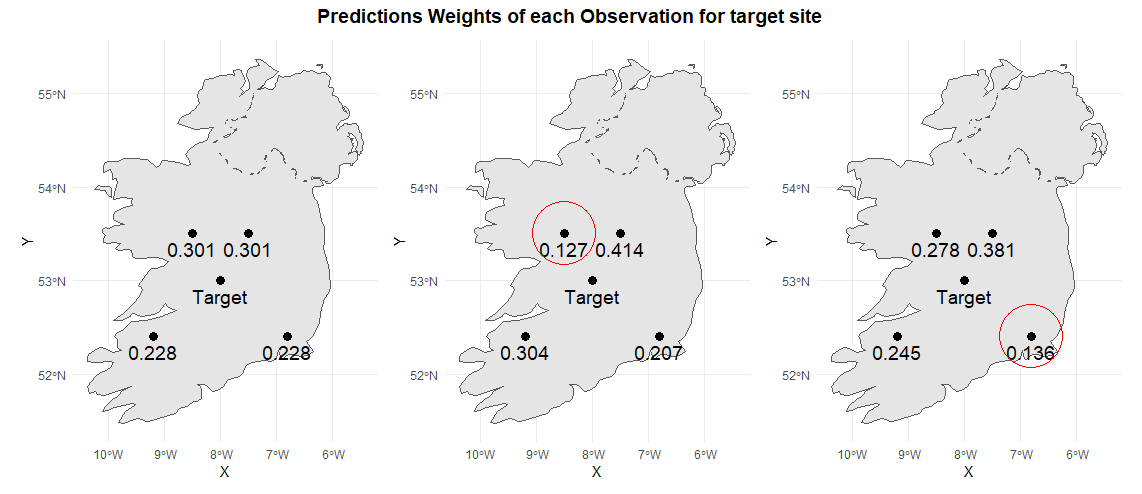}
\caption{\label{fig:Simulate}In  the first figure, a common nugget parameter of $\sigma = 0.3$ is used across all locations. In the second and third figure, $\sigma$ at the highlighted site is set to 0.6. This reduces to contribution of the perturbed site and increases the weighting of nearby sites.}
\end{figure}

\section{Simulation study parameter estimates.}
\label{app:Simulation_Tables}
\begin{table}[h!]
\centering
\caption{Estimated Parameter Values: Range and Spatial Standard Deviation}
\begin{tabular}{|c|c|c|c|c|c|c|c|c|c| }
\hline
\multirow{2}{*}{Noise Level ($\sigma$)} & \multicolumn{3}{c|}{Reliable Data Only} & \multicolumn{3}{c|}{All Data (Constant Params)} & \multicolumn{3}{c|}{All Data (Different Variance)} \\
\cline{2-10}
 & Range & $\sigma_{z}$ & $\rho$ & Range & $\sigma_{z}$ & $\rho$ & Range & $\sigma_{z}$ & $\rho$\\
\hline
0.5 & 240 & 0.74 &  & 204 & 0.63 &  & 237 & 0.70 & \\
\hline
\multicolumn{10}{|c|}{\textbf{AR(1) Process}} \\
\hline
0.5 & 226 & 0.76 & 0.80 & 234 & 0.79 & 0.87 & 224 & 0.80 & 0.85 \\
\hline
\end{tabular}
\label{tab:parameters}
\end{table}

\section{Modelling Data using Normal Transformation}
\label{app:N01_appoach}
An alternative model is transforming all stations observations to $\mathcal{N}(0,1)$, by first calculating the quantile at each station, and then using a standard normal CDF to transform the data to normal.  The spatio-temporal models are then performed using the transformed variable. When predictions are performed at an unobserved site, these can be converted back to wind speeds by first transforming back to a quantile using a standard normal CDF, and then transforming to a Weibull using an inverse Weibull CDF. For the inverse Weibull transform, we use the Weibull parameters estimated at each location during the bias correction process. We consider three modelling approaches, with the results detailed in Table \ref{tab:model_comparison_N01}.

\begin{table}[h]
\centering
\begin{tabular}{ |c|c|c| }  
 \hline
   \textbf{Model \#} & \textbf{Model Description} & \textbf{RMSE} \\
 \hline
 \multicolumn{3}{|c|}{\textbf{Independent Replicates}} \\
 \hline
 1 & Meteorological Stations only &  1.34   \\ 
 2 & Met and PWS (bias correction), single nugget variance  & 1.28  \\
 3 & Met and PWS (bias correction), group specific variance & 1.27  \\
\hline
\end{tabular}
\caption{Comparison of model performance using RMSE and CRPS.}
\label{tab:model_comparison_N01}
\end{table}

The RMSEs obtained are similar to those achieved using the square root transformation, and neither approach demonstrates a clear advantage for this dataset. We chose to focus on the square root transformation in the main text as it has been previously applied to Irish wind speed datasets. Additionally, it provides a more direct verification of the bias correction effect, as the $\mathcal{N}(0,1)$ transformation removes marginal distributional differences between stations, making it less informative for comparing pre- and post-bias correction performance.

\section{Parameter Confidence Intervals by Number of Timepoints}
\label{app:IntervalvsTime}

\begin{table}[ht]
\centering
\begin{tabular}{|c|ccc|ccc|}
\hline
\textbf{T} 
& \multicolumn{3}{c|}{\textbf{Effective Range}} 
& \multicolumn{3}{|c|}{\textbf{Spatial Standard Deviation}} \\
\cline{2-7}
& 2.5\% & 97.5\% & Interval Width 
& 2.5\% & 97.5\% & Interval Width \\
\hline
100   & 499 & 640 & 141 & 0.263 & 0.312 & 0.049 \\
500   & 450 & 500 &  50 & 0.284 & 0.304 & 0.020 \\
1000  & 442 & 475 &  33 & 0.310 & 0.325 & 0.015 \\
2000  & 453 & 476 &  23 & 0.311 & 0.321 & 0.010 \\
4392  & 445 & 471 &  26 & 0.335 & 0.343 & 0.008 \\
\hline
\end{tabular}
\caption{Posterior interval widths for selected parameters across increasing timepoints \( T \).}
\end{table}

\bibliographystyle{apalike}
\bibliography{main}

\end{document}